\documentclass[12pt,preprint]{aastex}

\shorttitle{Arecibo 6.7 GHz Methanol Maser Survey}

\begin{document}
\shortauthors{Pandian, Goldsmith, \& Deshpande}

\title{The Arecibo Methanol Maser Galactic Plane Survey--I: Data}

\author{Jagadheep D. Pandian \altaffilmark{1}, Paul F. Goldsmith \altaffilmark{2}, and Avinash A. Deshpande \altaffilmark{3,4}}
\altaffiltext{1}{Department of Astronomy, Cornell University, Ithaca, NY 14853; jagadheep@astro.cornell.edu}
\altaffiltext{2}{Jet Propulsion Laboratory, California Institute of Technology, Pasadena, CA 91109; Paul.F.Goldsmith@jpl.nasa.gov}
\altaffiltext{3}{Raman Research Institute, Bangalore 560 080, India; desh@rri.res.in}
\altaffiltext{4}{NAIC/Arecibo Observatory, HC3 Box 53995, Arecibo, PR 00612}

\begin{abstract}
We present the results of an unbiased survey for 6.7 GHz methanol masers in the Galactic plane carried out using the 305 m Arecibo radio telescope. A total of 18.2 square degrees was surveyed with uniform sampling at $35.2\degr \leq l \leq 53.7\degr$, $|b| \leq 0.41\degr$. The large collecting area of Arecibo and the sensitive C-Band High receiver allowed the survey to be complete at the level of 0.27 Jy making this the most sensitive blind survey carried out to date. We detected a total of 86 sources, 48 of which are new detections. Most of the new detections have a peak flux density below 2 Jy. Many methanol masers are clustered, reflecting the formation of massive stars in clusters.
\end{abstract}


\section{Introduction}
It is widely recognized that molecular masers can be a powerful tool in the study of early stages of star formation. The 6.7 GHz methanol maser, discovered by \citet{ment91}, is unique compared to its OH and H$_2$O counterparts in that it has been observed exclusively towards regions containing young, massive stars. Further, a number of recent studies suggest that 6.7 GHz methanol maser emission is associated with very early phases of massive star formation \citep{elli06, mini05, mini01}.

There have been approximately 519 methanol masers detected to date in various targeted and blind surveys \citep{pest05}. While many detections have been made by targeted surveys towards ultracompact H {\small II} regions selected based on their IRAS colors, blind surveys have shown that IRAS--based surveys underestimate the number of methanol masers by a factor of two \citep{elli96, szym02}. The total number of 6.7 GHz methanol masers in the Galaxy is thus estimated to be between 845 and 1200 \citep{van05}.

In this paper we present the results of a sensitive blind survey of a portion of the Galactic plane that is visible to the Arecibo\footnote{The Arecibo Observatory is part of the National Astronomy and Ionosphere Center, which is operated by Cornell University under a cooperative agreement with the National Science Foundation.} radio telescope. The data from such a survey can be productive for many purposes. Studies based on an unbiased sample of 6.7 GHz methanol masers are crucial for determining the link between methanol masers and massive star formation. It is of considerable interest to determine whether faint methanol masers discovered in a sensitive survey have different properties than those of bright masers; \citet{szym00b} discovered that methanol masers associated with IRAS sources that did not satisfy color criteria for ultracompact H {\small II} regions \citep{wood89, hugh89} were relatively weak. Finally, if all 6.7 GHz methanol masers are indeed associated with early phases of massive star formation, they are potential tools for studying Galactic spiral structure, and thus it is of interest to carry out surveys that sample as large a range in Galactic longitude as possible.

\section{Equipment}
Curiously, prior to 2004, no large radio telescope in U.S. was equipped with a receiver to detect the 6.7 GHz methanol line. In order to carry out this survey, we built a low noise cooled receiver for the Arecibo radio telescope for studying radio waves at frequencies between 6 and 8 GHz. The OrthoMode Transducer (OMT), which divides the incoming signal into two orthogonal polarizations, was designed based on the finline design of \citet{chat99}. It accepts two orthogonally linearly polarized inputs and sends each of them to a separate processing chain. The low noise amplifiers (LNA) use Microwave Monolithic Integrated Circuits (MMIC) with Indium Phosphide transistors to achieve a noise temperature of less than 4 K when cooled to 15 K. The LNAs were designed at Caltech \citep{wade05} using semiconductor chips fabricated by the Northcorp Grumman semiconductor foundry. The OMT and LNAs are cooled to 15 K to achieve low noise and to minimize the contribution of ohmic losses to the system temperature.

The feed horn is a wide angle corrugated horn designed to have an edge taper of -15 dB at 60\degr~off axis which marks the edge of the tertiary reflector of the Arecibo Gregorian feed system. The feed horn is kept at room temperature and is isolated from the cooled components by two thermal breaks, one between room temperature and 70 K, and another between 70 K and 15 K. This arrangement is used to minimize the heat load on the 15 K stage of the refrigerator. 

\begin{figure}
\plotone{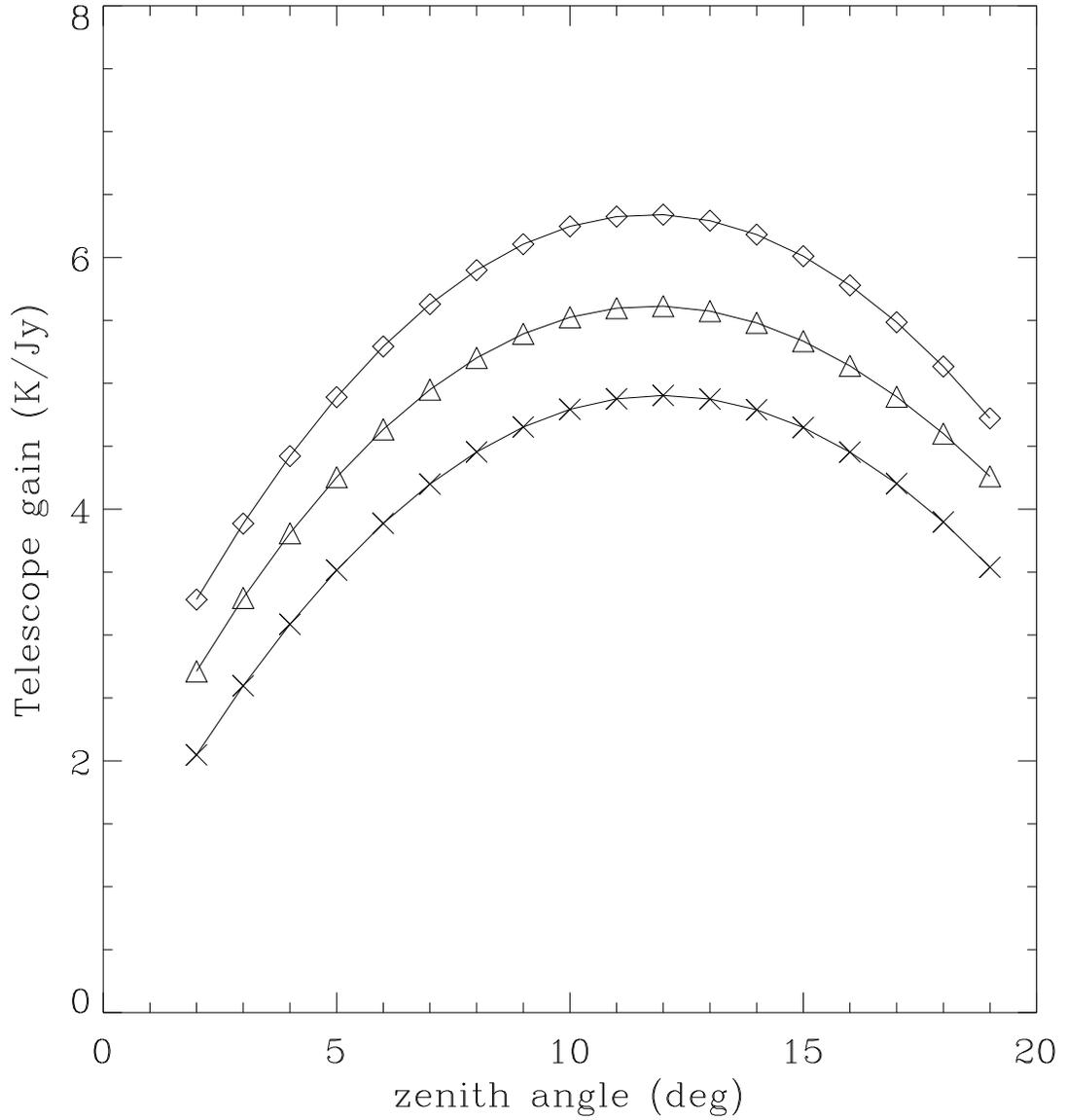}
\caption{The Arecibo telescope gain in K/Jy as a function of zenith angle for three different frequencies. The gain curves at 6600 MHz, 7000 MHz, and 7400 MHz are indicated by diamonds ($\Diamond$), triangles ($\triangle$), and crosses ($\times$) respectively.\label{gaincurve}}
\end{figure}

\begin{figure}
\includegraphics[angle=270]{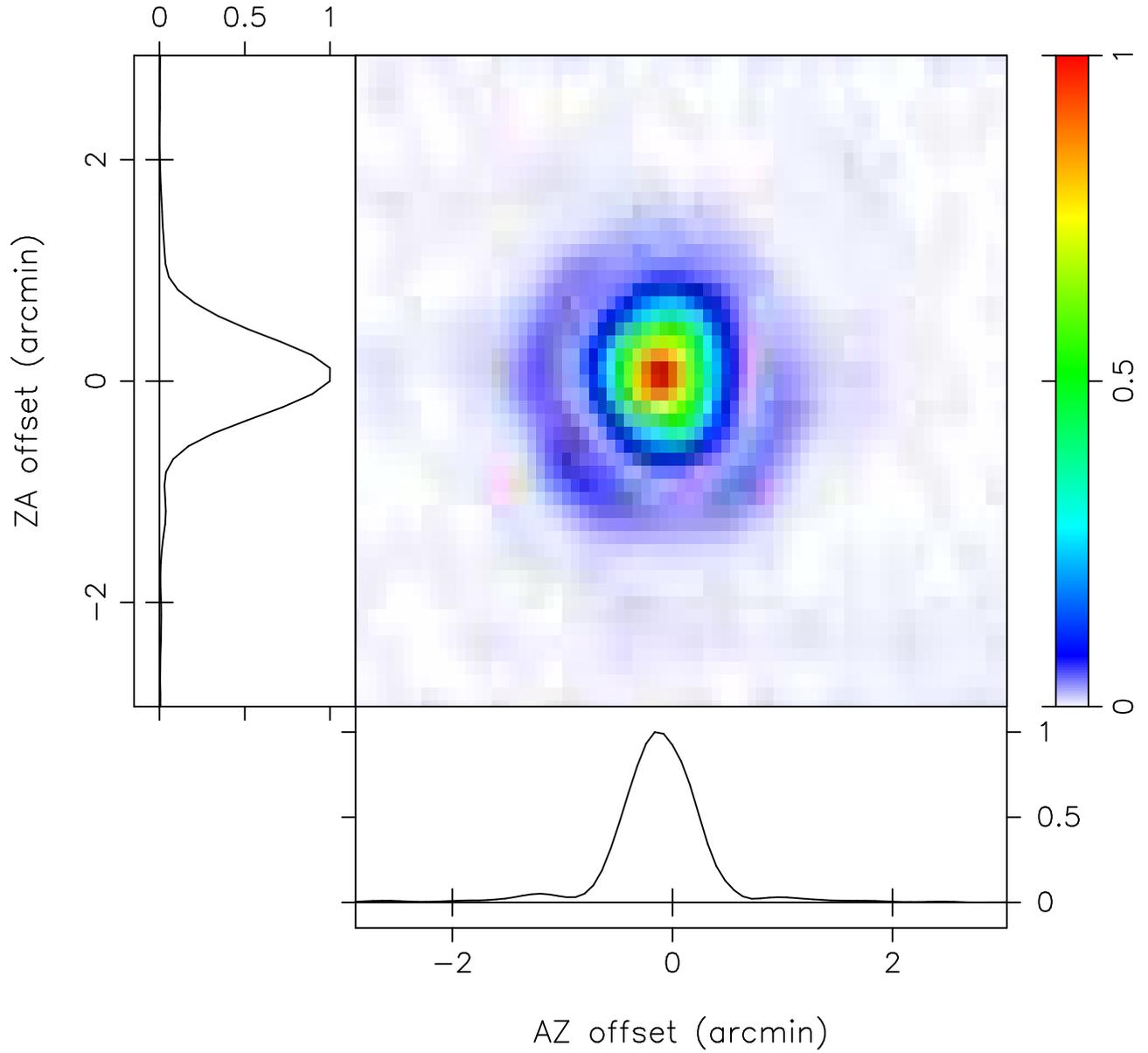}
\caption{Normalized beam response as a function of offsets in azimuth and zenith angle determined using full polarization measurements from a set of raster-scans on the continuum source B1040+123. The response shown here is that for the Stokes I parameter.\label{beam_map}}
\end{figure}

The receiver achieved a noise temperature of $9-11$ K across the 6 to 8 GHz band, which when combined with the stray radiation and spill-over contributions of the telescope and the sky emission resulted in a system temperature of $23-29$ K for one polarization and $28-34$ K for the other polarization. The telescope gain and system temperature are a function of the azimuth and zenith angle of the telescope with the system temperature increasing (and the telescope gain decreasing) at high zenith angles. The gain curve of the telescope as a function of zenith angle is shown in Figure~\ref{gaincurve}, and the beam pattern at 6.6 GHz is shown in Figure~\ref{beam_map}. The elliptical shape of the main beam is a result of the area of the primary reflector illuminated by the Arecibo Gregorian feed system \citep{gold96}.

A more detailed description of the receiver design and performance can be found in \citet{pand06a}.

\section{Observations}
The observations were made between June 2004 and March 2006 using the 6 -- 8 GHz (C-Band High) receiver at the 305 m Arecibo radio telescope. The telescope has an rms pointing error of 7\arcsec~and an average full width half maximum (FWHM) beamwidth of 40\arcsec. The backend was the digital ``interim'' correlator which has four boards, each of which can be configured independently. The boards were configured into two pairs, each pair recording the autocorrelations of the two orthogonal linear polarizations with 9-level sampling, and 2048 channels per polarization to cover a bandwidth of 3.125 MHz. The two pairs of boards were staggered in frequency in order to provide a net velocity coverage of +110 to -70 km s$^{-1}$, with a velocity resolution of 0.14 km s$^{-1}$ after Hanning smoothing at the 6668.518 MHz rest frequency of the $5_1-6_0$ A$^+$ transition of methanol (\citealt{ment91}; a more accurate measurement of the rest frequency is presented by \citealt{brec95} as 6668.519 MHz; this results in a change in velocities by 0.04 km s$^{-1}$, which is less than the precision of the velocity of peak emission given in Table 1). The observations were centered on a frequency corresponding to the local standard of rest. 

The nominal target region of the sky was $35\degr \la l \la 55\degr$, $|b| \la 0.4\degr$, which we divided into 100 sub-regions. Each sub-region was mapped in a rectangular grid (in right ascension and declination), the separation between grid points being 15\arcsec~in right ascension and 20\arcsec~in declination. Overall, we sampled approximately 780,000 sky positions. A noise diode with known noise temperature was fired every 250--270 sec in order to calibrate the antenna temperature scale of the data. The data were calibrated in terms of flux density by using the elevation--dependent gain curves (see Figure~\ref{gaincurve}).   Since the telescope gain drops at both high and low zenith angles, we planned the survey such that all observations would be carried out when the target region was between zenith angles of 5\degr and 18\degr. The residuals obtained from the fit to the telescope gain curves indicate that the flux calibration has an uncertainty of approximately 10\%. 

Part of the observations were done in ``drift mode'' where the telescope is fixed and the sky drifts through the beam. The integration time per grid point in this mode was 1 sec, which yielded an rms noise level of $\sim$ 70 mJy in each spectral channel after Hanning smoothing averaging both polarizations.  
Between June and November 2004, the primary, secondary, and tertiary reflector surfaces were readjusted, resulting in improved telescope performance. Following this, we carried out the rest of the survey by driving the telescope at twice the sidereal rate and integrating for 0.5 sec on each sample of the sky. This resulted in an rms noise level of $\sim$ 85 mJy in each spectral channel after Hanning smoothing and averaging both polarizations.  Positions observed during bad weather were typically re-observed so that the spatial noise distribution in a map would be as uniform as possible.

\begin{figure}
\plotone{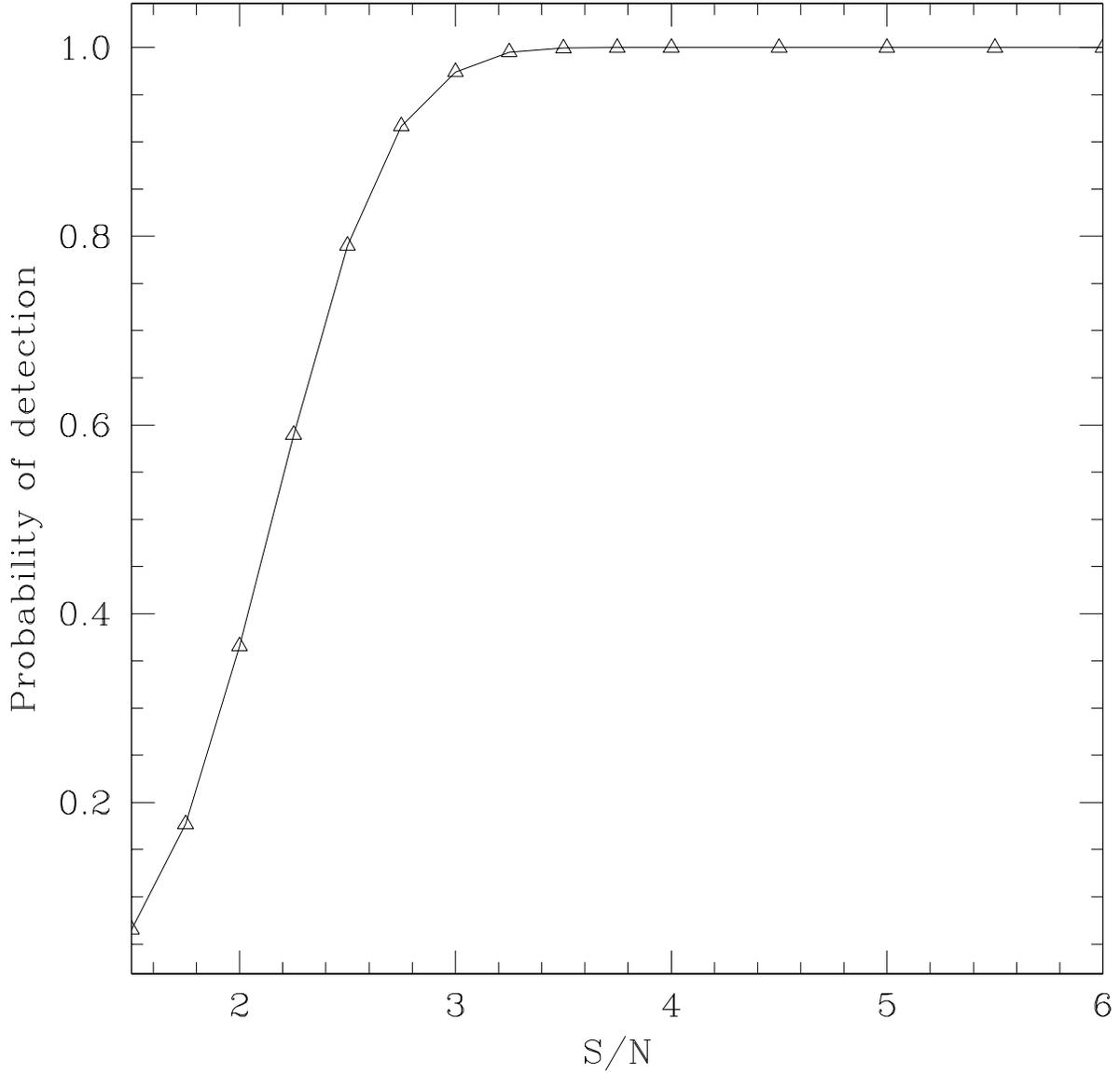}
\caption{The probability of detection of a signal as a function of signal to noise ratio for our source extraction algorithm determined using simulated data cubes. The signal detection probability is greater than 95\% for signal to noise ratio of 3.0. The detection threshold used results in a false alarm probability of less than $10^{-7}$.\label{probdetect}}
\end{figure}

The data reduction was carried out in IDL using procedures maintained by the observatory. The data for each sub-region was reduced to give a data cube. We implemented an automated matched filter algorithm in three dimensions (two spatial dimensions and one frequency dimension) to extract candidate sources from the data cube. The data were first convolved in the frequency domain with a Gaussian of a given line width, since Gaussians are good models for describing maser line profiles. Each spectral channel image was then output as a FITS file, from which point sources were extracted using SExtractor \citep{bert96}. A convolution filter was applied to the images within SExtractor before extracting the point sources. In our case, this filter was set to the point spread function of the telescope (Figure \ref{beam_map}) since the masers are unresolved by single dish telescopes.  The angular response of the telescope was approximated by a single Gaussian fitted to the two dimensional main lobe of the telescope beam. The procedure above was repeated for different Gaussian line widths in the frequency domain since maser line widths are variable. The convolving line widths varied from 0.05 km s$^{-1}$s to 0.5 km s$^{-1}$.  We tested the algorithm on simulated data cubes to determine the probability of detection for different signal to noise ratios. The results of the simulation are shown in Figure \ref{probdetect}. The simulation shows that our source extraction algorithm achieves greater than 95\% probability of detection for signals that have signal to noise ratio higher than 3.0, and since the noise in the data cubes is no greater than 0.09 Jy, we conclude that our source catalog is complete at the level of 0.27 Jy.

The candidate sources obtained above were re-observed using a one minute position switched observation. These observations were centered on the velocity of the source, and had one pair of boards with a velocity resolution of 0.14 km s$^{-1}$, and a second pair of boards with a velocity resolution of 0.03 km s$^{-1}$. These observations enabled us to distinguish real sources from statistical noise fluctuations, and in addition gave high velocity resolution, high signal to noise spectra for all sources. Sources weaker than $\sim 0.5$ Jy were typically observed with a two minutes on source to obtain a high signal to noise ratio. Since we average the two polarizations, our results are not affected by the small levels of linear polarization as observed by \citet{elli02}.

There were instances where emission was discovered at the edges of our data cubes. In these cases, we made additional observations around these positions to determine the spatial peak of emission before carrying out follow-up observations.

\section{Results}

We detected a total of 86 sources, 48 of which are new detections. The region $35.2\degr \leq l \leq 53.7\degr$, $|b| \leq 0.41\degr$ was completely sampled, while a larger area of $34.7\degr \leq l \leq 54.3\degr$, $|b| \leq 0.54\degr$ was incompletely sampled. The region surveyed is shown in Figure \ref{mappedregion} along with the positions of the detected sources. The J2000 positions, velocity range of maser emission, $v_{min},~v_{max}$, velocity of peak emission, $v_p$, peak flux density, $S_p$, integrated flux density, $S_i$, epoch of observation, and the reference of the discovery for each source are given in Table \ref{tbl1}. Individual spectra for all sources are shown in Figure \ref{maserspectra}. Spectra of previously detected sources are included in part because most have much higher signal to noise compared to previous observations, and in part for variability studies.

We note that the source 49.57--0.38 listed in the general catalog of 6.7 GHz methanol masers \citep{pest05} is a sidelobe of source 49.49--0.39 (W51) as mentioned by \citet{szym00a}. We were also unable to detect the source 49.66--0.45 which was observed towards IRAS 19220+1431 by \citet{slys99}. While this could be due to source variability, the non-detection of this source by \citet{van95}, coincidence of the velocity of peak emission with that of the strong nearby source 49.49--0.39, and the difference between the velocity range of the maser and that of the CS (2--1) line \citep{bron96} strongly suggest that this source is also a sidelobe of 49.49--0.39.

Previously published data on known detections is summarized in Table \ref{tbl2}. In particular, we list the difference between the position determined from our data cubes and that published previously, $\Delta \theta$, peak flux density, $S_p$ and epoch of the observation. We found that the positions of the IRAS sources surveyed by \citet{szym00a}, that are listed in the general catalog of methanol masers \citep{pest05} are incorrect, often by 1\arcmin. This in turn results in incorrect Galactic coordinates being calculated, with which the sources are named. Further, since the true position of the methanol maser is often offset from the position of the IRAS source, we choose to re-label the names of these sources with Galactic coordinates calculated from our survey. We also re-label a source in the general catalog that was detected in the unpublished Onsala Survey, whose coordinates differ from ours by 6\arcmin. We retain the names of all other sources that were published elsewhere, even if their Galactic coordinates differ from that calculated from our positions, for the sake of consistency.

\section{Notes on selected sources}
{\noindent \bf 38.92-0.36.} This source was detected by \citet{szym00a} towards IRAS 19012+0505 as a 5.4 Jy source, but was undetected by \citet{szym02}. We measured a peak flux density of 1.26 Jy towards this source in Sep. 2005. Thus, peak flux density of this source has decreased by a factor of 4.3 in 6.5 years, and by at least a factor of 2 in one year given the detection limits of \citet{szym02}.

{\noindent \bf 41.87-0.10.} This is a mysterious source. We detected this source at R.A. $19^h08^m10^s.8$, Dec. 07\degr54\arcmin04\arcsec\ in the survey data taken on Oct. 1, 2005. The source had a peak flux density of 0.84 Jy at a velocity of 23.3 km s$^{-1}$ (Figure \ref{mystery}). However, the follow-up data taken on Oct. 19, 2005 revealed a different spectrum with a peak flux density of 0.22 Jy at a velocity of 15.8 km s$^{-1}$, while the feature around 23.3 km s$^{-1}$ had a peak flux density of only 0.09 Jy (Figure \ref{mystery}). Future attempts to detect this source were unsuccessful. The $3\sigma$ limits on the source are 0.27 Jy, 0.08 Jy and 0.07 Jy on Oct. 22, 2005, Nov. 15, 2005 and Mar. 23, 2006 respectively. Since there are no nearby strong sources at this velocity, it is unlikely that this source is a side-lobe detection. Since the source happened to lie at the edge of our map of the region, we are not sure about the accuracy of the source position.

{\noindent \bf W49N region.} This is a complex region containing a cluster of methanol masers. In order to distinguish the different masing regions, we first identified the different spectral features at various locations in the region. Next, we identified the spatial peaks of each spectral feature. This enabled us to identify at least five distinct regions with methanol emission. Four of these regions were also identified by \citet{casw95}.

The source 43.15+0.02 has a peak flux density of 24.3 Jy at 13.3 km s$^{-1}$. 43.16+0.02 has a main double peaked feature between 8 and 10 km s$^{-1}$, and other features between 15 and 22 km s$^{-1}$. The source 43.17+0.01 has features between 18 and 23 km s$^{-1}$, which are blended with 43.16+0.02. 43.17-0.00 has a 2.7 Jy peak around --1 km s$^{-1}$, while 43.18-0.01 has a 0.9 Jy peak around 11 km s$^{-1}$. In addition, there is an additional center that emits at 13.7 km s$^{-1}$, the position of which is not well determined from our data cubes, but is around R.A. $19^h10^m12^s.6$, Dec. 09\degr06\arcmin11\arcsec. Unfortunately, we do not have follow-up data at this position for verification, and consequently, we are not including this source in our source list. An image made from aperture synthesis will be required to define all the methanol emitting regions unambiguously.

Since the sources 43.16+0.02 and 43.17+0.01 have blended features, we carried out a simple deconvolution procedure to get the real spectra. If two sources, labeled 1 and 2 have spectra $S_1^a(\nu)$ and $S_2^a(\nu)$ respectively, the measured spectra $S_1^m(\nu)$ and $S_2^m(\nu)$ can be written as 
\begin{eqnarray}
S_1^m(\nu) & = & S_1^a(\nu) + \alpha S_2^a(\nu) \\
S_2^m(\nu) & = & S_2^a(\nu) + \alpha S_1^a(\nu) 
\end{eqnarray}
Here, the parameter $\alpha$ determines the amount of blending, and can in theory be determined from the beam shape and the separation of the two sources. We assume a symmetric pattern for the main beam, so that the amount of contamination of source 1 onto source 2 is the same as the blending of source 2 onto source 1. The equations above can be inverted easily to obtain
\begin{eqnarray}
S_1^a(\nu) & = & \frac{S_1^m(\nu) - \alpha S_2^m(\nu)}{1 - \alpha^2} \\
S_2^a(\nu) & = & \frac{S_2^m(\nu) - \alpha S_1^m(\nu)}{1 - \alpha^2} 
\end{eqnarray}

In the case of sources 43.16+0.02 and 43.17+0.01, the spectral features between 8 and 10 km s$^{-1}$ are associated only with 43.16+0.02. We used this fact to determine the parameter $\alpha$ to be 0.15. The deconvolution explained above was used to obtain the spectra shown in Figure \ref{maserspectra}.

{\noindent \bf W51 region.} W51 is another complex region that has a cluster of methanol masers. One of the masers in this complex is very strong ($> 700$ Jy), which makes the data analysis for this region unusually difficult due to very relative ease with which this source can be detected in sidelobes of the antenna response pattern. We identified four distinct sites of methanol emission using techniques explained in the W49N section above. We also observed that sources 49.49-0.39 and 49.49-0.37 were blended between 57 and 60 km s$^{-1}$. We deconvolved these two sources using the technique explained above. The parameter $\alpha$ in the equations above was determined to be 0.023 using the feature at 52 km s$^{-1}$, which is associated only with the source 49.49-0.39. The deconvolved spectra are shown in Figure \ref{maserspectra}.

\section{Discussion}

This paper presents the basic results of the Arecibo Methanol Maser Galactic Plane Survey (AMGPS). An analysis of the distribution of sources, association with massive star formation and implications for Galactic structure will be given in a separate paper \citep{pand06b}.

A part of our target region ($l \leq 40\degr$) was covered by the blind survey of \citet{szym02} which had a 1$\sigma$ sensitivity of 0.4 to 0.7 Jy. We detected 21 new sources in this region. The histogram of flux densities of the sources detected in our survey is shown in Figure \ref{fluxdist}. From the figure, it is clear that a significant fraction of new detections have flux densities less than or equal to 2 Jy. This highlights the importance of high sensitivity in carrying out unbiased surveys of this type. 

It is also evident from Figure \ref{fluxdist} that there are some new detections that are reasonably strong, with one source having a peak flux density of 11.3 Jy. These sources are in regions that have not been covered by unbiased surveys. Since they have not been detected in previous targeted surveys, it is very likely that they do not have infrared counterparts in the IRAS catalog, nor do they have previously detected OH maser emission associated with them.  The latter is probably a consequence of the fact that OH maser emission is typically a factor of 7 weaker than that of the 6.7 GHz methanol emission \citep{casw95}.

\begin{figure}
\plotone{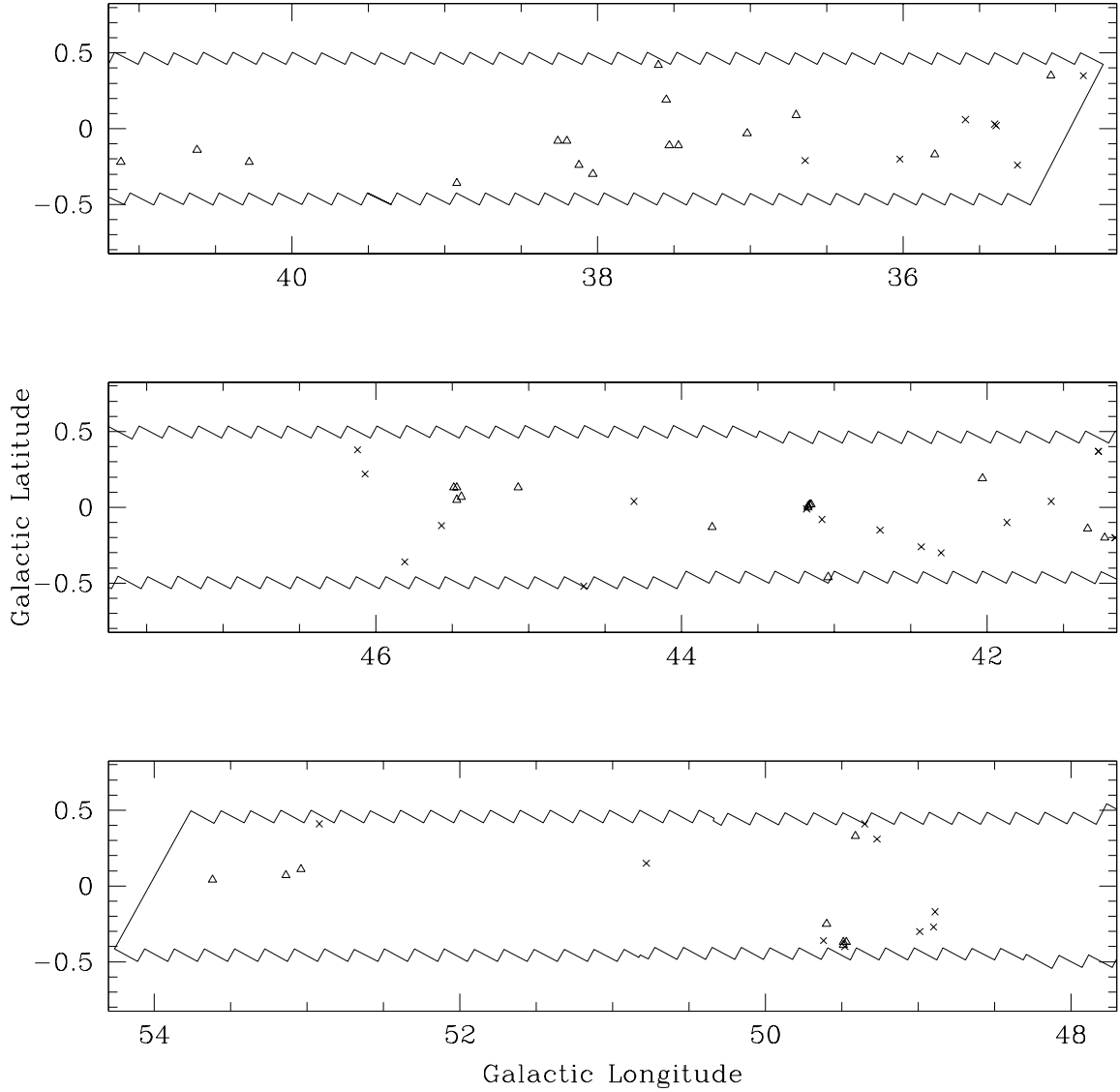}
\caption{The region mapped in the Arecibo Methanol Maser Galactic Plane Survey shown in Galactic coordinates. New detections are denoted by crosses and prior detections are denoted by open triangles. The cluster of sources around $l=43.2\degr$ belongs to the W49 region, while the cluster around $l=49.5\degr$ belongs to the W51 region.\label{mappedregion}}
\end{figure}

\begin{figure}
\plotone{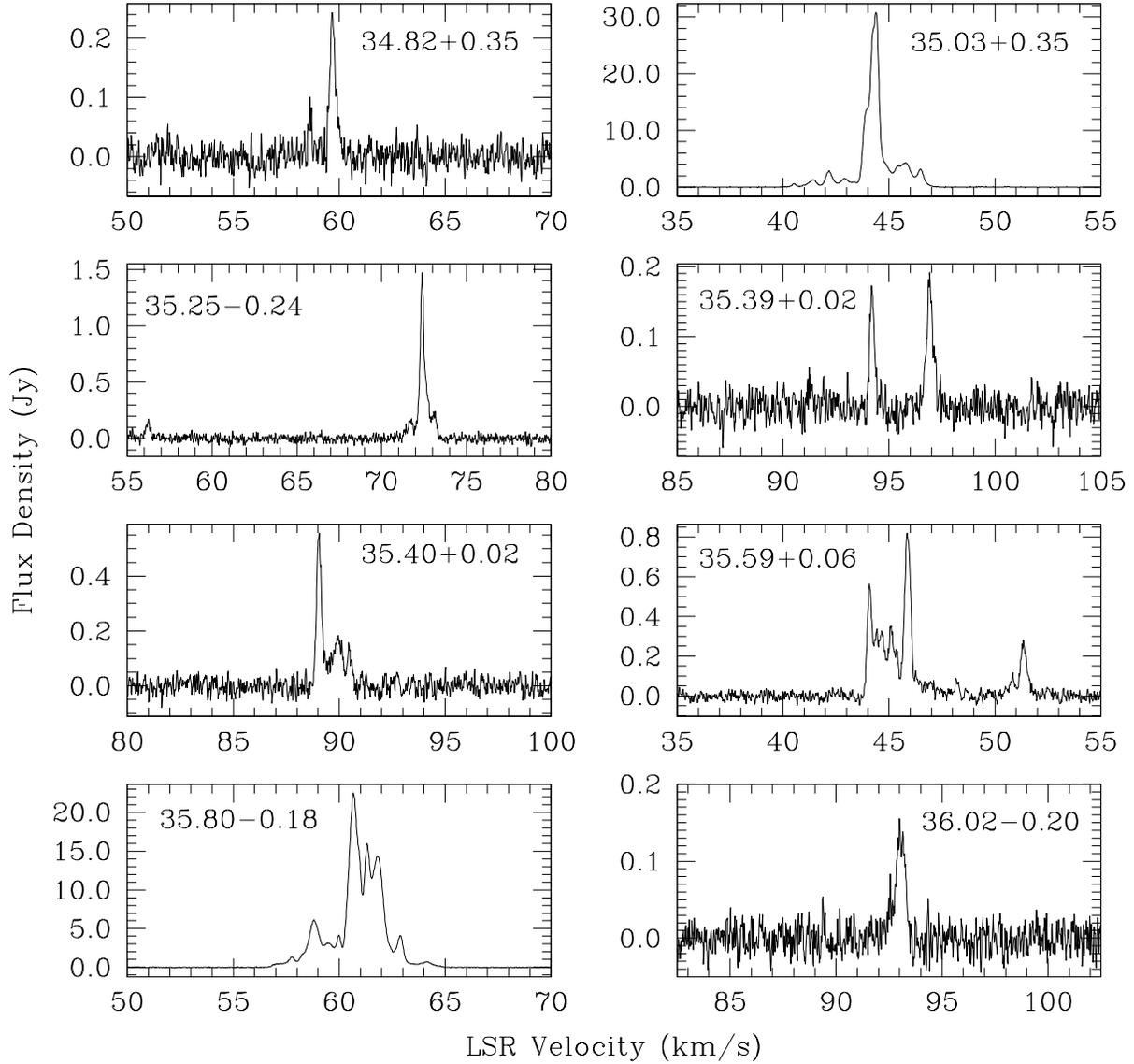}
\caption{Spectra of 6.7 GHz methanol masers detected in the Arecibo Methanol Maser Galactic Plane Survey. The velocity resolution of all spectra is 0.03 km s$^{-1}$.}\label{maserspectra}
\end{figure}

\begin{figure}
\plotone{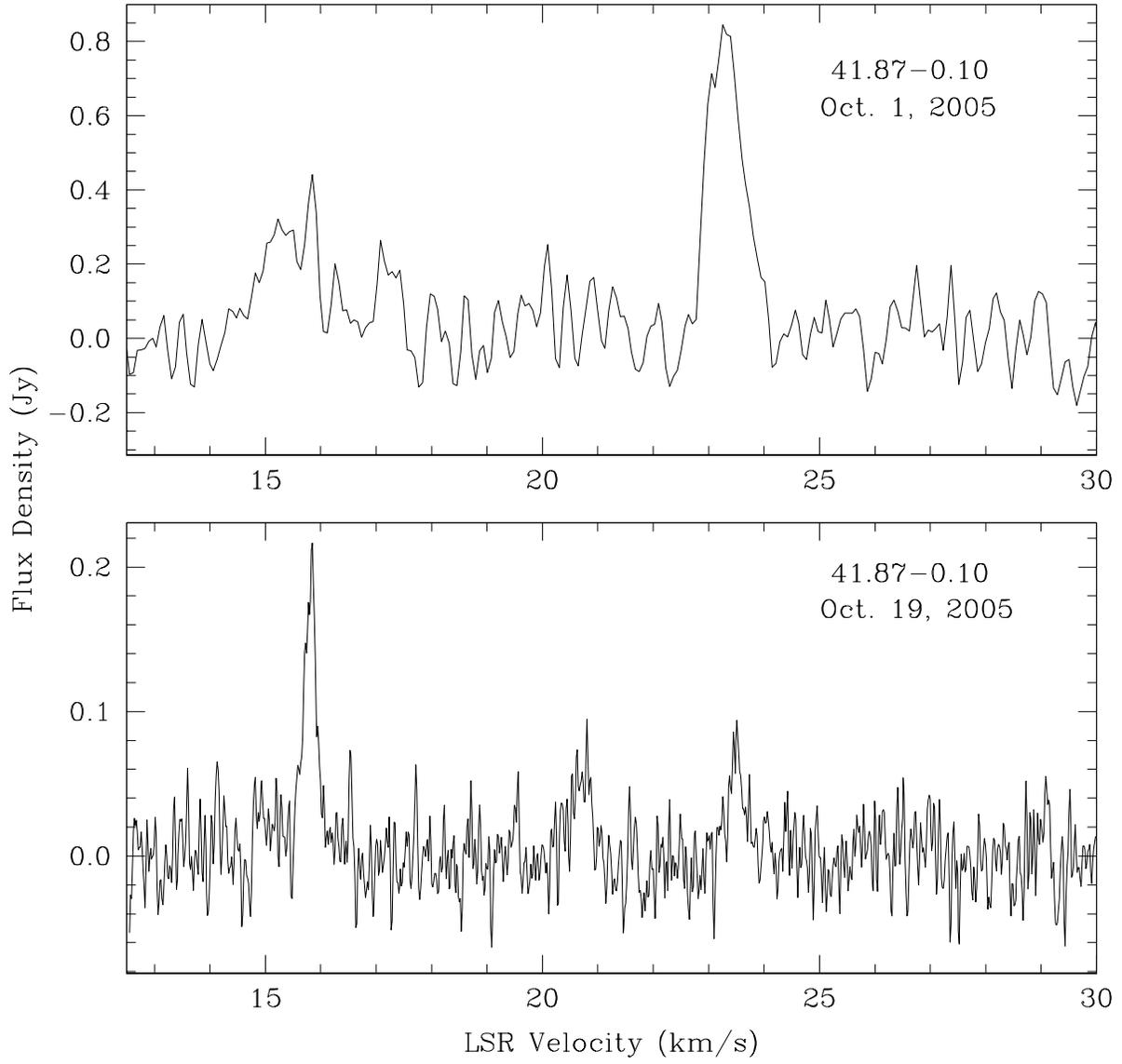}
\caption{The top panel shows the spectrum of source 41.87--0.10 from the survey data obtained on Oct. 1, 2005. The bottom panel shows the spectrum of the same source obtained in follow-up observations on Oct. 19, 2005. The source was undetected in all subsequent observations. \label{mystery}}
\end{figure}

\begin{figure}
\plotone{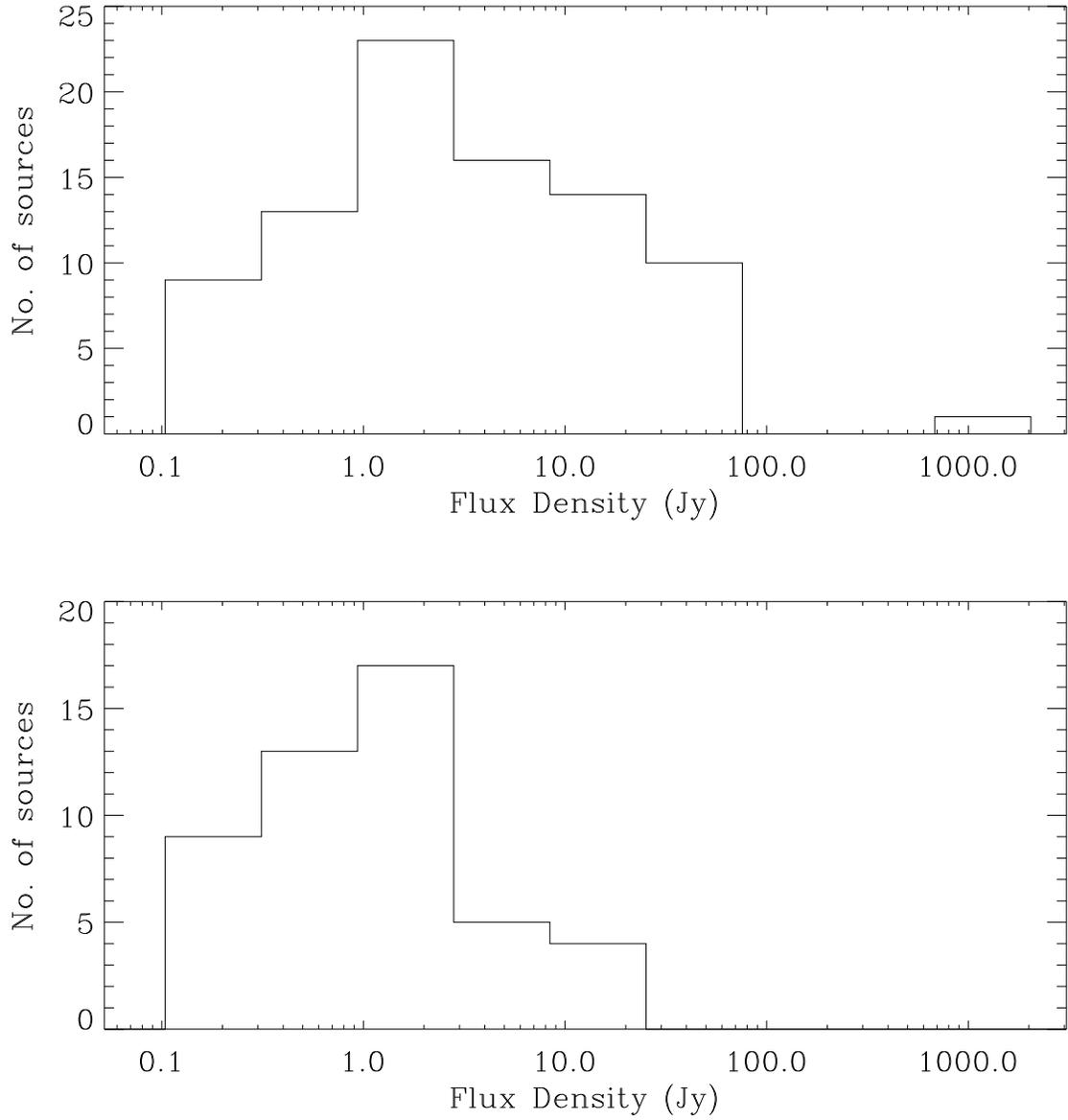}
\caption{The distribution of flux densities of methanol masers discovered in the AMGPS. The top panel shows the histogram for all sources detected in the survey, while the bottom panel shows the histogram for new detections. Note that the lowest bin will be affected by incompleteness.\label{fluxdist}}
\end{figure}

It can be seen from Table \ref{tbl1} and Figure \ref{mappedregion} that there are very few sources at $l > 49.5\degr$, which corresponds to the tangent point of the Carina-Sagittarius spiral arm. Since the number density of methanol masers as a function of Galactic longitude drops significantly beyond the tangent point of a spiral arm, surveys towards the inner regions of the Galaxy ($l \la 35\degr$) are likely to yield many more sources than surveys towards other regions. Another feature that is evident from Figure \ref{mappedregion} is the dearth of sources between Galactic longitudes of 46\degr and 49\degr. A similar patchiness in the maser distribution was found in the earlier blind survey of \citet{elli96}, as exemplified by the region $325\degr \leq l \leq 326.4\degr$.  Our result is perhaps more dramatic, especially considering the higher sensitivity of AGMPS.  We do not have at this time any explanation for this patchy distribution.

Further, it is apparent from Table \ref{tbl1} that many methanol masers occur in clusters. Notable examples other than the W49N and W51 regions are the groups around $(l,~b)$ = (37.7--38.2, -0.2) (there seem to be two groups of masers in this region), (41.1, -0.2) and (45.5, 0.10). There are several other pairs of masers that are located close to each other both spatially and in velocity. We take this to reflect the clustered nature of massive star formation.

The data in Table \ref{tbl2} can be used to estimate variability in methanol masers. Since this is not a systematic study of variability like the work of \citet{goed04}, we cannot estimate any quantitative measure for variability. Qualitatively, we can determine variability for a maser if the change in its measured peak flux density exceeds the combined measurement errors. To do this, we first determined the errors in the flux densities for each survey by combining the accuracy of the flux calibration with the error resulting from antenna pointing errors. A pair of measurements differing by more than their error estimates is interpreted to be suggestive of variability. This does not take into account variability or lack of it among individual velocity features of the masers. Using this metric, we found that 28 out of the 38 masers listed in Table \ref{tbl2} are likely to be variable. These peak flux densities of these sources have typically changed by over 25\%.  The fraction of the maser population that is variable is comparable to that determined by \citet{casw95} (sources marked as ``v'' and ``sv'' in Table 1 of \citealt{casw95}).

\acknowledgments

We are grateful to Phil Perillat for many discussions regarding the performance of the telescope, and routines for basic data reduction.  We thank German Cortes--Medellin and Lynn Baker for their contributions to the receiver system development, Kurt Kabelac and David Overbaugh for machining most of the dewar components, and Rajagopalan Ganesan and Lisa Locke for calibrating the receiver. We are also grateful to numerous other members of staff at the Arecibo Observatory who helped us in the receiver installation and calibration, and in setting up our observations. We also thank the referee for a number of constructive comments that improved the paper. This work was supported in part by the Jet Propulsion Laboratory, California Institute of Technology. This research has made use of NASA's Astrophysics Data System.

\clearpage

\begin{deluxetable}{ccccccccc}
\tabletypesize{\footnotesize}
\tablecaption{6.7 GHz methanol masers discovered in the Arecibo Methanol Maser Galactic Plane Survey (AMGPS).\label{tbl1}}
\tablewidth{0pt}
\tablehead{
\colhead{Source ($l$, $b$)} & \colhead{$\alpha$ (J2000)} & \colhead{$\delta$ (J2000)} & \colhead{$\Delta v$} & \colhead{$v_p$} & \colhead{$S_p$} & \colhead{$S_i$} & \colhead{Epoch} & \colhead{Ref.} \\
\colhead{} & \colhead{} & \colhead{} & \colhead{(km s$^{-1}$)} & \colhead{(km s$^{-1}$)} & \colhead{(Jy)} & \colhead{(Jy km s$^{-1}$)} & \colhead{} & \colhead{det.} 
}
\startdata
34.82+0.35 & 18 53 37.4 & 01 50 32 & 58.5, 60.1 & 59.7 & 0.24 & 0.10 & Oct. 2004 & 1 \\
35.03+0.35 & 18 54 01.3 & 02 01 28 & 40.2, 47.4 & 44.4 & 30.77 & 27.41 & Oct. 2004 & 2 \\
35.25--0.24 & 18 56 30.9 & 01 57 11 & 56.0, 73.3 & 72.4 & 1.47 & 0.71 & Oct. 2004 & 1 \\
35.39+0.02 & 18 55 51.2 & 02 11 37 & 94.0, 97.2 & 96.9 & 0.19 & 0.11 & Oct. 2004 & 1 \\
35.40+0.03 & 18 55 51.1 & 02 12 25 & 88.8, 90.7 & 89.1 & 0.56 & 0.30 & Oct. 2004 & 1 \\
35.59+0.06 & 18 56 04.3 & 02 23 28 & 43.8, 51.8 & 45.9 & 0.82 & 0.96 & Oct. 2004 & 1 \\
35.79--0.17 & 18 57 17.5 & 02 28 04 & 56.6, 64.9 & 60.7 & 22.49 & 37.74 & Oct. 2004 & 7 \\
36.02--0.20 & 18 57 45.8 & 02 39 15 & 92.4, 93.5 & 93.0 & 0.16 & 0.08 & Oct. 2004 & 1 \\
36.64--0.21 & 18 58 55.9 & 03 12 05 & 77.0, 79.3 & 77.3 & 1.64 & 0.52 & Apr. 2005 & 1 \\
36.70+0.09 & 18 57 59.3 & 03 24 05 & 52.2, 63.2 & 54.7 & 7.00 & 8.51 & Apr. 2005 & 7 \\
36.84--0.02 & 18 59 39.3 & 03 27 55 & 52.8, 64.2 & 61.7 & 1.66 & 4.65 & Sep. 2005 & 1 \\
36.90--0.41 & 19 00 08.6 & 03 20 35 & 83.1, 85.1 & 84.7 & 0.45 & 0.26 & Apr. 2005 & 1 \\
36.92+0.48 & 18 57 00.7 & 03 46 01 & --36.3, --35.6 & --35.9 & 1.52 & 0.51 & Sep. 2005 & 1 \\
37.02--0.03 & 18 59 05.1 & 03 37 47 & 77.5, 85.3 & 78.4 & 7.24 & 4.77 & Sep. 2005 & 7 \\
37.38--0.09 & 18 59 52.3 & 03 55 12 & 67.5, 70.9 & 70.6 & 0.19 & 0.13 & Sep. 2005 & 1 \\
37.47--0.11 & 19 00 08.1 & 03 59 49 & 53.6, 63.3 & 54.7 & 9.83 & 22.41 & Sep. 2005 & 7 \\
37.53--0.11 & 19 00 17.4 & 04 03 15 & 48.2, 56.6 & 50.0 & 4.48 & 5.77 & Sep. 2005 & 5 \\
37.55+0.19 & 18 59 11.4 & 04 12 14 & 78.1, 88.2 & 83.7 & 5.27 & 6.44 & Sep. 2005 & 4 \\
37.60+0.42 & 18 58 28.0 & 04 20 46 & 84.6, 94.7 & 85.8 & 17.30 & 25.29 & Sep. 2005 & 7 \\
37.74--0.12 & 19 00 38.0 & 04 13 18 & 49.9, 50.5 & 50.3 & 0.93 & 0.25 & Sep. 2005 & 1 \\
37.76--0.19 & 19 00 56.5 & 04 12 08 & 54.9, 66.0 & 55.1 & 0.59 & 1.46 & Sep. 2005 & 1 \\
37.77--0.22 & 19 01 03.0 & 04 12 13 & 68.8, 70.3 & 69.6 & 0.82 & 0.37 & Sep. 2005 & 1 \\
38.03--0.30 & 19 01 51.4 & 04 24 16 & 54.6, 65.9 & 55.6 & 11.71 & 18.10 & Sep. 2005 & 7 \\
38.08--0.27 & 19 01 48.4 & 04 27 25 & 66.7, 67.8 & 67.5 & 0.59 & 0.18 & Sep. 2005 & 1 \\
38.12--0.24 & 19 01 45.2 & 04 30 32 & 66.5, 79.7 & 70.2 & 1.92 & 5.67 & Sep. 2005 & 5 \\
38.20--0.08 & 19 01 20.1 & 04 39 37 & 77.7, 88.5 & 79.6 & 11.09 & 16.26 & Sep. 2005 & 7 \\
38.26--0.08 & 19 01 27.2 & 04 42 09 & 6.1, 15.9 & 15.4 & 7.03 & 5.39 & Sep. 2005 & 7 \\
38.26--0.20 & 19 01 54.0 & 04 38 38 & 64.1, 73.5 & 70.2 & 0.72 & 1.28 & Sep. 2005 & 1 \\
38.56+0.15 & 19 01 10.3 & 05 04 26 & 23.1, 31.2 & 31.5 & 0.18 & 0.13 & Sep. 2005 & 1 \\
38.60--0.21 & 19 02 34.0 & 04 56 40 & 61.4, 69.5 & 62.6 & 0.57 & 0.59 & Sep. 2005 & 1 \\
38.66+0.08 & 19 01 36.7 & 05 07 42 & --31.9, --30.7 & --31.5 & 2.19 & 0.71 & Sep. 2005 & 1 \\
38.92--0.36\tablenotemark{a} & 19 03 39.7 & 05 09 36 & 30.8, 33.5 & 31.9 & 1.26 & 1.29 & Sep. 2005 & 5 \\
39.39--0.14 & 19 03 45.3 & 05 40 39 & 58.2, 75.5 & 60.4 & 1.14 & 0.77 & Sep. 2005 & 1 \\
39.54--0.38 & 19 04 53.5 & 05 41 59 & 47.4, 49.4 & 47.8 & 0.20 & 0.24 & Sep. 2005 & 1 \\
40.28--0.22\tablenotemark{a} & 19 05 42.1 & 06 26 08 & 62.4, 85.7 & 73.9 & 24.47 & 68.03 & Oct. 2005 & 5 \\
40.62--0.14 & 19 06 02.3 & 06 46 37 & 29.7, 36.7 & 31.1 & 15.20 & 6.97 & Oct. 2005 & 2 \\
40.94--0.04 & 19 06 16.1 & 07 06 00 & 36.2, 43.2 & 36.6 & 2.35 & 1.50 & Oct. 2005 & 1 \\
41.08--0.13 & 19 06 49.3 & 07 11 01 & 57.2, 58.4 & 57.5 & 0.79 & 0.33 & Oct. 2005 & 1 \\
41.12--0.11 & 19 06 50.7 & 07 13 57 & 33.1, 37.4 & 36.6 & 1.14 & 0.63 & Oct. 2005 & 1 \\
41.12--0.22\tablenotemark{a} & 19 07 15.4 & 07 10 54 & 55.0, 66.6 & 63.4 & 2.01 & 1.23 & Oct. 2005 & 5 \\
41.16--0.20 & 19 07 15.1 & 07 13 20 & 61.6, 63.8 & 63.6 & 0.27 & 0.20 & Oct. 2005 & 1 \\
41.23--0.20\tablenotemark{a} & 19 07 21.9 & 07 17 06 & 54.0, 64.9 & 55.4 & 3.46 & 7.55 & Oct. 2005 & 5 \\
41.27+0.37 & 19 05 24.6 & 07 35 02 & 19.4, 20.6 & 20.3 & 0.26 & 0.16 & Oct. 2005 & 1\\
41.34--0.14 & 19 07 22.8 & 07 25 17 & 6.6, 15.0 & 11.7 & 17.80 & 25.40 & Oct. 2005 & 6 \\
41.58+0.04 & 19 07 09.4 & 07 42 19 & 10.4, 12.3 & 11.9 & 0.50 & 0.22 & Oct. 2005 & 1 \\
41.87--0.10\tablenotemark{b} & 19 08 10.8 & 07 54 04 & 15.5, 23.7 & 15.8 & 0.22 & 0.10 & Oct. 2005 & 1 \\
42.03+0.19\tablenotemark{c} & 19 07 29.0 & 08 10 39 & 6.8, 17.3 & 12.8 & 26.32 & 30.10 & Oct. 2005 & 6 \\
42.30--0.30 & 19 09 44.2 & 08 11 33 & 26.2, 34.7 & 28.1 & 6.33 & 5.07 & Oct. 2005 & 1 \\
42.43--0.26 & 19 09 50.2 & 08 19 32 & 65.8, 69.1 & 66.8 & 1.91 & 1.61 & Oct. 2005 & 1 \\
42.70--0.15 & 19 09 55.8 & 08 36 56 & --47.1, --39.0 & --42.9 & 3.25 & 4.03 & Oct. 2005 & 1 \\
43.04--0.46\tablenotemark{a} & 19 11 39.7 & 08 46 32 & 54.1, 63.6 & 54.8 & 7.19 & 10.39 & Oct. 2005 & 5 \\
43.08--0.08 & 19 10 22.4 & 08 59 01 & 9.6, 14.9 & 10.2 & 9.20 & 4.11 & Oct. 2005 & 1 \\
43.15+0.02 & 19 10 11.9 & 09 05 24 & 12.3, 14.3 & 13.3 & 24.29 & 12.03 & Oct. 2005 & 3 \\
43.16+0.02 & 19 10 13.9 & 09 06 16 & 6.8, 22.1 & 9.3 & 27.77 & 50.54 & Oct. 2005 & 3 \\
43.17+0.01 & 19 10 16.1 & 09 06 16 & 18.1, 22.4 & 19.0 & 9.27 & 15.05 & Oct. 2005 & 3 \\
43.17--0.00 & 19 10 17.7 & 09 05 54 & --1.7, 4.2 & --1.2 & 2.68 & 1.44 & Oct. 2005 & 3 \\
43.18--0.01 & 19 10 20.2 & 09 06 06 & 10.3, 11.6 & 11.1 & 0.88 & 0.63 & Oct. 2005 & 1 \\
43.80--0.13 & 19 11 54.8 & 09 35 48 & 38.4, 43.6 & 39.6 & 55.56 & 43.18 & Mar. 2006 & 2 \\
44.31+0.04 & 19 12 16.4 & 10 07 44 & 55.0, 56.6 & 55.7 & 0.66 & 0.43 & Mar. 2006 & 1 \\
44.64--0.52 & 19 14 54.6 & 10 10 02 & 48.8, 49.9 & 49.3 & 0.55 & 0.29 & Mar. 2006 & 1 \\
45.07+0.13 & 19 13 22.5 & 10 51 01 & 56.8, 60.0 & 57.8 & 48.27 & 22.26 & Mar. 2006 & 2 \\
45.44+0.07 & 19 14 19.0 & 11 09 07 & 49.1, 50.6 & 50.0 & 1.09 & 0.71 & Mar. 2006 & 3 \\
45.47+0.05 & 19 14 24.5 & 11 09 40 & 55.4, 59.8 & 56.0 & 5.69 & 9.40 & Mar. 2006 & 2 \\
45.47+0.13 & 19 14 08.2 & 11 12 24 & 57.0 73.5 & 65.7 & 5.49 & 4.88 & Mar. 2006 & 2 \\
45.49+0.13 & 19 14 11.8 & 11 13 13 & 56.7, 66.4 & 57.2 & 8.77 & 4.98 & Mar. 2006 & 3 \\
45.57--0.12 & 19 15 13.2 & 11 10 25 & 1.2, 9.8 & 1.6 & 0.40 & 0.30 & Mar. 2006 & 1 \\
45.81--0.36 & 19 16 31.9 & 11 16 22 & 54.7, 70.8 & 59.9 & 11.31 & 8.41 & Mar. 2006 & 1 \\
46.07+0.22 & 19 14 55.6 & 11 46 12 & 22.3, 25.1 & 23.3 & 1.23 & 1.20 & Apr. 2006 & 1 \\
46.12+0.38 & 19 14 26.4 & 11 53 24 & 57.5, 62.9 & 59.0 & 1.13 & 1.05 & Apr. 2006 & 1 \\
48.89--0.17 & 19 21 47.5 & 14 04 58 & 57.2, 57.5 & 57.3 & 0.13 & 0.02 & Oct. 2004 & 1 \\
48.90--0.27 & 19 22 10.1 & 14 02 38 & 63.6, 72.5 & 72.0 & 0.83 & 0.57 & Oct. 2004 & 1 \\
48.99--0.30 & 19 22 26.3 & 14 06 37 & 62.5, 72.6 & 71.6 & 0.58 & 0.55 & Oct. 2004 & 1 \\
49.27+0.31 & 19 20 44.8 & 14 38 29 & --6.8, 7.5 & --3.2 & 8.12 & 13.71 & Oct. 2004 & 1 \\
49.35+0.41 & 19 20 33.2 & 14 45 48 & 66.1, 69.2 & 68.0 & 7.00 & 6.72 & Apr. 2005 & 1 \\
49.41+0.33\tablenotemark{a} & 19 20 58.9 & 14 46 46 & --27.0, --9.8 & --12.1 & 9.25 & 23.18 & Apr. 2005 & 5 \\
49.47--0.37 & 19 23 38.3 & 14 29 58 & 55.6, 76.1 & 63.8 & 7.01 & 17.35 & Oct. 2004 & 3 \\
49.48--0.40 & 19 23 46.1 & 14 29 38 & 47.7, 65.3 & 51.4 & 8.06 & 14.90 & Oct. 2004 & 1 \\
49.49--0.37 & 19 23 40.0 & 14 31 04 & 54.4, 65.7 & 56.1 & 32.27 & 49.27 & Oct. 2004 & 2 \\
49.49--0.39 & 19 23 43.9 & 14 30 31 & 49.9, 65.0 & 59.3 & 738.4 & 574.7 & Oct. 2004 & 3 \\
49.60--0.25\tablenotemark{a} & 19 23 26.7 & 14 40 19 & 59.9, 66.7 & 62.9 & 38.11 & 78.16 & Oct. 2004 & 5 \\
49.62--0.36 & 19 23 52.8 & 14 38 10 & 48.8, 60.1 & 49.3 & 1.23 & 1.13 & Oct. 2004 & 1 \\
50.78+0.15 & 19 24 17.2 & 15 53 54 & 47.6, 50.9 & 49.1 & 5.26 & 2.79 & Sep. 2005 & 1 \\
52.92+0.41 & 19 27 35.2 & 17 54 26 & 38.8, 45.0 & 39.1 & 6.64 & 4.68 & Mar. 2006 & 1 \\
53.04+0.11\tablenotemark{a} & 19 28 55.7 & 17 52 01 & 9.7, 10.5 & 10.1 & 1.66 & 0.69 & Mar. 2006 & 5 \\
53.14+0.07\tablenotemark{a} & 19 29 17.5 & 17 56 24 & 23.4, 25.4 & 24.6 & 1.02 & 0.54 & Mar. 2006 & 5 \\
53.62+0.04\tablenotemark{a} & 19 30 22.6 & 18 20 28 & 18.2, 19.5 & 19.0 & 18.94 & 7.16 & Mar. 2006 & 5 \\
\enddata
\tablecomments{Each source is defined by its Galactic coordinates, and is followed by its equatorial coordinates.  $\Delta v$ is the velocity range of emission, $v_p$ is the velocity of peak emission, $S_p$ is the peak flux density, and $S_i$ is the integrated flux density. Epoch refers to the date of observation.  The final column gives reference for detection of each source; (1) refers to first detection in the AMGPS, while other numbers indicate detections in prior surveys indicated, in addition to the present work.}
\tablenotetext{a}{This source is identified by different $l$, $b$ coordinates in \citet{pest05} due to incorrect coordinates being quoted for the IRAS source towards which \citet{szym00a} detected methanol maser emission. We re-label this source using our coordinates which are more accurate for the methanol maser than the coordinates of the IRAS source.}
\tablenotetext{b}{This source was detected at the edge of a data cube, and could not be detected in successive observations of the region. Hence, the actual coordinates of this source could be different from those quoted here.}
\tablenotetext{c}{The coordinates of this source detected in the Onsala Survey (which is not yet published in a refereed journal), and listed in \citet{pest05} are very different from our coordinates. We think that this could be due to a software problem, similar to the one giving rise to incorrect coordinates for IRAS sources published in \citet{pest05}. Hence, we label this source with $l$, $b$ determined from our coordinates.}
\tablerefs{
(1) This paper; (2) Menten 1991; (3) Caswell et al. 1995; (4) Slysh et al. 1999; (5) Szymczak et al. 2000; (6) Pestalozzi et al. 2002; (7) Szymczak et al. 2002.}
\end{deluxetable}

\begin{deluxetable}{ccccccc}
\tabletypesize{\footnotesize}
\tablecaption{Previously published data is compared with our data for known detections.\label{tbl2}}
\tablewidth{0pt}
\tablehead{
\colhead{Source ($l$, $b$)} & \colhead{$\alpha$ (J2000)} & \colhead{$\delta$ (J2000)} & \colhead{$\Delta \theta$} & \colhead{Ref.} & \colhead{Epoch} & \colhead{$S_p$} \\
\colhead{} & \colhead{} & \colhead{} & \colhead{(\arcsec)} & \colhead{} & \colhead{} & \colhead{(Jy)}
}
\startdata
35.03+0.35 & 18 54 01.3 & 02 01 28 & \nodata & 1 & Oct. 2004 & 30.8 \\
           & 18 54 00.6 & 02 00 50 & 39 & 7 & 2000 & 38.9 \\
           & 18 54 00.6 & 02 01 15 & 17 & 3 & 1992 & 56 \\
           & 18 54 01.8 & 02 01 19 & 12 & 2 & Jun. 1991 & 50 \\[1.5mm]
35.79--0.17 & 18 57 17.5 & 02 28 04 & \nodata & 1 & Oct. 2004 & 22.5 \\
            & 18 57 16.1 & 02 27 44 & 29 & 7 & 2000 & 24.5 \\[1.5mm]
36.70+0.09 & 18 57 59.3 & 03 24 05 & \nodata & 1 & Apr. 2005 & 7.0 \\
           & 18 58 00.9 & 03 23 30 & 42 & 7 & 2000 & 8.6 \\[1.5mm]
37.02--0.03 & 18 59 05.1 & 03 37 47 & \nodata & 1 & Sep. 2005 & 7.2 \\
            & 18 58 59.9 & 03 37 40 & 78 & 7 & 2000 & 7.1 \\[1.5mm]
37.47--0.11 & 19 00 08.1 & 03 59 49 & \nodata & 1 & Sep. 2005 & 9.8 \\
            & 19 00 06.7 & 03 59 27 & 30 & 7 & 2000 & 12.5 \\[1.5mm]
37.53--0.11 & 19 00 17.4 & 04 03 15 & \nodata & 1 & Sep. 2005 & 4.5 \\
            & 19 00 14.4 & 04 02 35 & 60 & 7 & 2000 & 5.8 \\
            & 19 00 15.8 & 04 03 07 & 25 & 5 & Jul. 1999 & 5.7 \\[1.5mm]
37.55+0.19 & 18 59 11.4 & 04 12 14 & \nodata & 1 & Sep. 2005 & 5.3 \\
           & 18 59 11.6 & 04 12 08 & 7 & 7 & 2000 & 8.1 \\
           & 18 59 09.9 & 04 12 14 & 22 & 5 & Aug. 1999 & 6.6 \\
           & 18 59 09.9 & 04 12 14 & 22 & 4 & Mar./Apr. 1999 & 4 \\[1.5mm]
37.60+0.42 & 18 58 28.0 & 04 20 46 & \nodata & 1 & Sep. 2005 & 17.3 \\
           & 18 58 28.5 & 04 20 34 & 14 & 7 & 2000 & 24.3 \\[1.5mm]
38.03--0.30 & 19 01 51.4 & 04 24 16 & \nodata & 1 & Sep. 2005 & 11.7 \\
            & 19 01 50.0 & 04 23 54 & 30 & 7 & 2000 & 18.6 \\[1.5mm]
38.12--0.24 & 19 01 45.2 & 04 30 32 & \nodata & 1 & Sep. 2005 & 1.9 \\
            & 19 01 47.6 & 04 30 32 & 36 & 7 & 2000 & 4.2 \\
            & 19 01 43.0 & 04 30 45 & 35 & 5 & Jul. 1999 & 2.3 \\[1.5mm]
38.20--0.08 & 19 01 20.1 & 04 39 37 & \nodata & 1 & Sep. 2005 & 11.1 \\
            & 19 01 22.8 & 04 39 10 & 49 & 7 & 2000 & 8.4 \\[1.5mm]
38.26--0.08 & 19 01 27.2 & 04 42 09 & \nodata & 1 & Sep. 2005 & 7.0 \\
            & 19 01 28.7 & 04 42 02 & 23 & 7 & 2000 & 7.9 \\[1.5mm]
38.92--0.36\tablenotemark{a} & 19 03 39.7 & 05 09 36 & \nodata & 1 & Sep. 2005 & 1.3 \\
            & 19 03 43.5 & 05 09 49 & 58 & 5 & Mar. 1999 & 5.4 \\[1.5mm]
40.28--0.22\tablenotemark{a} & 19 05 42.1 & 06 26 08 & \nodata & 1 & Oct. 2005 & 24.5 \\
           & 19 05 36.4 & 06 26 09 & 85 & 5 & Aug. 1999 & 18 \\[1.5mm]
40.62--0.14 & 19 06 02.3 & 06 46 37 & \nodata & 1 & Oct. 2005 & 15.2 \\
            & 19 06 00.8 & 06 46 37 & 22 & 3 & 1992 & 17 \\
            & 19 06 01.1 & 06 46 35 & 18 & 2 & Jun. 1991 & 14 \\[1.5mm]
41.12--0.22\tablenotemark{a} & 19 07 15.4 & 07 10 54 & \nodata & 1 & Oct. 2005 & 2.0 \\
            & 19 07 16.9 & 07 10 02 & 57 & 5 & Aug. 1999 & 3.7 \\[1.5mm]
41.23--0.20\tablenotemark{a} & 19 07 21.9 & 07 17 06 & \nodata & 1 & Oct. 2005 & 3.5 \\
            & 19 07 21.0 & 07 17 25 & 23 & 5 & Aug. 1999 & 3.2 \\[1.5mm]
41.34--0.14 & 19 07 22.8 & 07 25 17 & \nodata & 1 & Oct. 2005 & 17.8 \\
            & 19 07 21.870 & 07 25 17.34 & 14 & 6 & ? & 51 \\[1.5mm]
42.03+0.19\tablenotemark{b} & 19 07 29.0 & 08 10 39 & \nodata & 1 & Oct. 2005 & 26.3 \\
           & 19 07 20.8 & 08 14 12 & 245 & 6 & ? & 12 \\[1.5mm]
43.04--0.46\tablenotemark{a} & 19 11 39.7 & 08 46 32 & \nodata & 1 & Oct. 2005 & 7.2 \\
            & 19 11 37.4 & 08 46 30 & 34 & 5 & Jun. 1999 & 10 \\[1.5mm]
43.15+0.02 & 19 10 11.9 & 09 05 24 & \nodata & 1 & Oct. 2005 & 24.3 \\
           & 19 10 10.6 & 09 05 27 & 19 & 3 & 1992 & 26 \\[1.5mm]
43.16+0.02 & 19 10 13.9 & 09 06 16 & \nodata & 1 & Oct. 2005 & 27.8 \\
           & 19 10 15.3 & 09 06 08 & 22 & 5 & Jun. 1999 & 31 \\
           & 19 10 11.9 & 09 06 15 & 30 & 3 & 1992 & 22 \\
           & 19 10 13.5 & 09 06 11 & 8 & 2 & Jun. 1991 & 28 \\[1.5mm]
43.17+0.01 & 19 10 16.1 & 09 06 16 & \nodata & 1 & Oct. 2005 & 9.4 \\
           & 19 10 14.6 & 09 06 18 & 22 & 3 & 1992 & 12.8 \\[1.5mm]
43.17--0.00 & 19 10 17.7 & 09 05 54 & \nodata & 1 & Oct. 2005 & 2.7 \\
            & 19 10 16.2 & 09 06 02 & 24 & 3 & 1992 & 3.4 \\[1.5mm]
43.80--0.13 & 19 11 54.8 & 09 35 48 & \nodata & 1 & Mar. 2006 & 55.6 \\
            & 19 11 53.3 & 09 35 46 & 22 & 5 & Jun. 1999 & 79 \\
            & 19 11 53.8 & 09 35 46 & 15 & 3 & 1992 & 144 \\
            & 19 11 53.9 & 09 35 51 & 14 & 2 & Jun. 1991 & 152 \\[1.5mm]
45.07+0.13 & 19 13 22.5 & 10 51 01 & \nodata & 1 & Mar. 2006 & 48.3 \\
           & 19 13 22.0 & 10 50 52 & 12 & 5 & Jul. 1999 & 45 \\
           & 19 13 22.2 & 10 50 52 & 10 & 3 & 1992 & 33 \\
           & 19 13 22.1 & 10 50 54 & 9 & 2 & Jun. 1991 & 42 \\[1.5mm]
45.44+0.07 & 19 14 19.0 & 11 09 07 & \nodata & 1 & Mar. 2006 & 1.1 \\
           & 19 14 21.7 & 11 09 14 & 40 & 5 & May 1999 & 4.1 \\
           & 19 14 17.9 & 11 08 58 & 19 & 3 & 1992 & 1.9 \\[1.5mm]
45.47+0.05 & 19 14 24.5 & 11 09 40 & \nodata & 1 & Mar. 2006 & 5.7 \\
           & 19 14 25.7 & 11 09 26 & 23 & 2 & Jun. 1991 & 3 \\[1.5mm]
45.47+0.13 & 19 14 08.2 & 11 12 24 & \nodata & 1 & Mar. 2006 & 5.5 \\
           & 19 14 08.6 & 11 12 27 & 7 & 5 & Jul. 1999 & 10 \\
           & 19 14 07.5 & 11 12 17 & 12 & 3 & 1992 & 5.3 \\
           & 19 14 08.8 & 11 12 28 & 10 & 2 & Jun. 1991 & 15 \\[1.5mm]
45.49+0.13 & 19 14 11.8 & 11 13 13 & \nodata & 1 & Mar. 2006 & 8.8 \\
           & 19 14 11.3 & 11 13 07 & 9 & 3 & 1992 & 13.4 \\[1.5mm]
49.41+0.33\tablenotemark{a} & 19 20 58.9 & 14 46 46 & \nodata & 1 & Apr. 2005 & 9.3 \\
           & 19 20 57.0 & 14 46 40 & 28 & 5 & Jul. 1999 & 13 \\[1.5mm]
49.47--0.37 & 19 23 38.3 & 14 28 58 & \nodata & 1 & Oct. 2004 & 7.0 \\
            & 19 23 38.3 & 14 30 09 & 71 & 3 & 1992 & 12 \\[1.5mm]
49.49--0.37 & 19 23 40.0 & 14 31 04 & \nodata & 1 & Oct. 2004 & 32.3 \\
            & 19 23 39.5 & 14 31 01 & 8 & 3 & 1992 & 33 \\[1.5mm]
49.49--0.39 & 19 23 43.9 & 14 30 31 & \nodata & 1 & Oct. 2004 & 738.4 \\
            & 19 23 43.4 & 14 30 35 & 8 & 3 & 1992 & 850 \\
            & 19 23 42.1 & 14 30 41 & 28 & 2 & Jun. 1991 & 979 \\[1.5mm]
49.60--0.25\tablenotemark{a} & 19 23 26.7 & 14 40 19 & \nodata & 1 & Oct. 2004 & 38.1 \\
            & 19 23 27.9 & 14 38 11 & 129 & 5 & Jul. 1999 & 29 \\[1.5mm]
53.04+0.11\tablenotemark{a} & 19 28 55.7 & 17 52 01 & \nodata & 1 & Mar. 2006 & 1.7 \\
           & 19 28 54.0 & 17 51 56 & 25 & 5 & Jul. 1999 & 3.3 \\[1.5mm]
53.14+0.07\tablenotemark{a} & 19 29 17.5 & 17 56 24 & \nodata & 1 & Mar. 2006 & 1.0 \\
           & 19 29 17.7 & 17 56 20 & 5 & 5 & Aug. 1999 & 1.1 \\[1.5mm]
53.62+0.04\tablenotemark{a} & 19 30 22.6 & 18 20 28 & \nodata & 1 & Mar. 2006 & 18.9 \\
           & 19 30 28.1 & 18 20 53 & 82 & 5 & Jul. 1999 & 6.3 \\
\enddata
\tablecomments{Some sources have coordinates substantially different from values determined from our survey. The two--dimensional angular offset between the two positions, $\Delta \theta$,  is given in arcseconds. For each source the first line gives our results , while earlier detections are given by the following lines.  Many sources have significantly different peak flux densities, $S_p$, indicating variability. The epochs for previous observations are given in order to establish a timescale for variability.}
\tablenotetext{a}{This source is identified by different $l$, $b$ coordinates in \citet{pest05} due to incorrect coordinates being quoted for the IRAS source towards which \citet{szym00a} detected methanol maser emission. We re-label this source after its correct coordinates.}
\tablenotetext{b}{The coordinates of this source detected in the Onsala Survey (which is not yet published in a refereed journal), and listed in \citet{pest05} are very different from our coordinates. We think that this could be due to a software problem, similar to the one giving rise to incorrect coordinates for IRAS sources published in \citet{pest05}. Hence, we label this source with $l$, $b$ determined from our coordinates.}
\tablerefs{
(1) This paper; (2) Menten 1991; (3) Caswell et al. 1995; (4) Slysh et al. 1999; (5) Szymczak et al. 2000; (6) Pestalozzi et al. 2002; (7) Szymczak et al. 2002.}
\end{deluxetable}

\clearpage

\clearpage
{\plotone{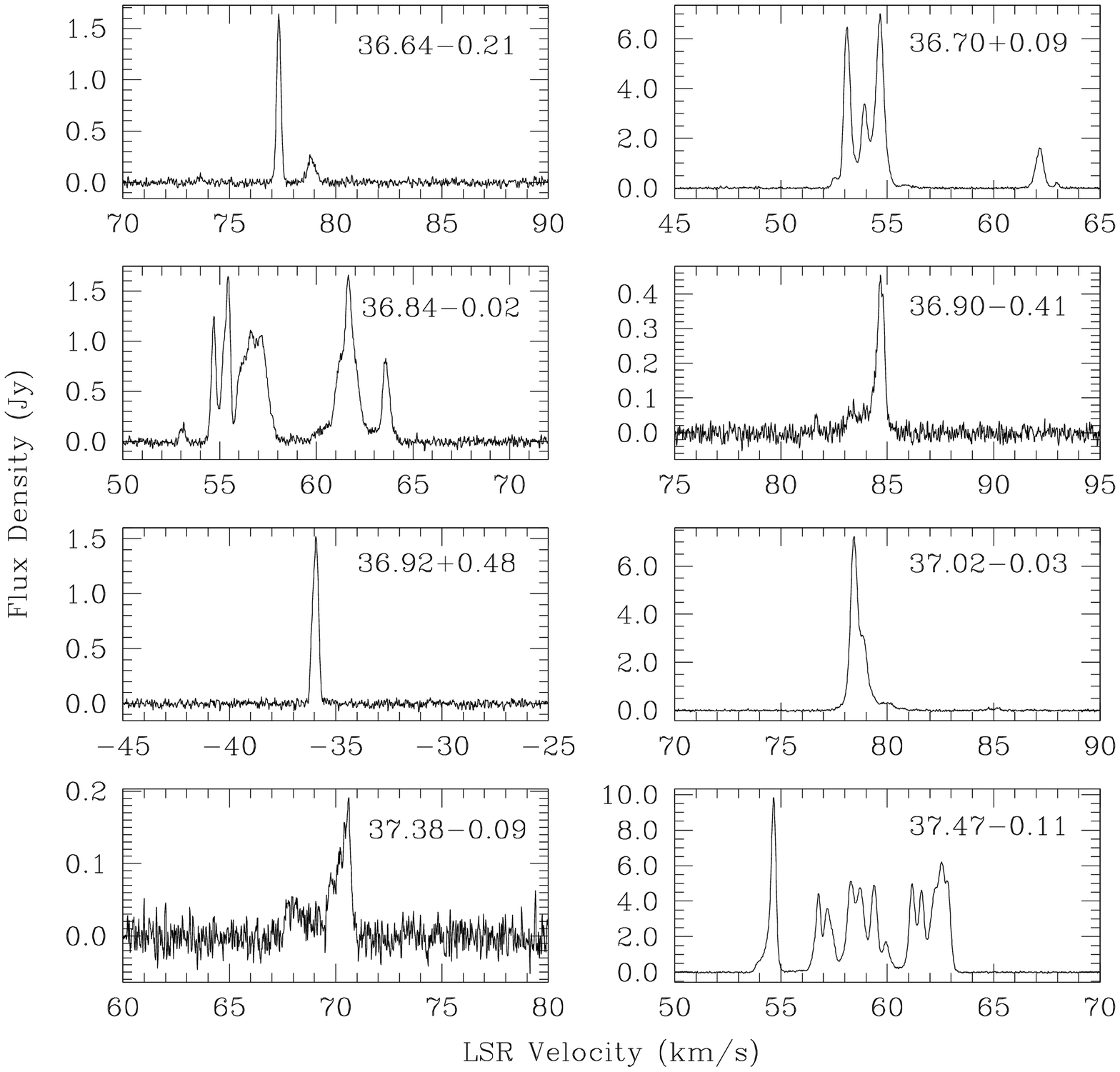}}\\[5mm]
\centerline{Fig. 6. --- Continued.}
\clearpage
{\plotone{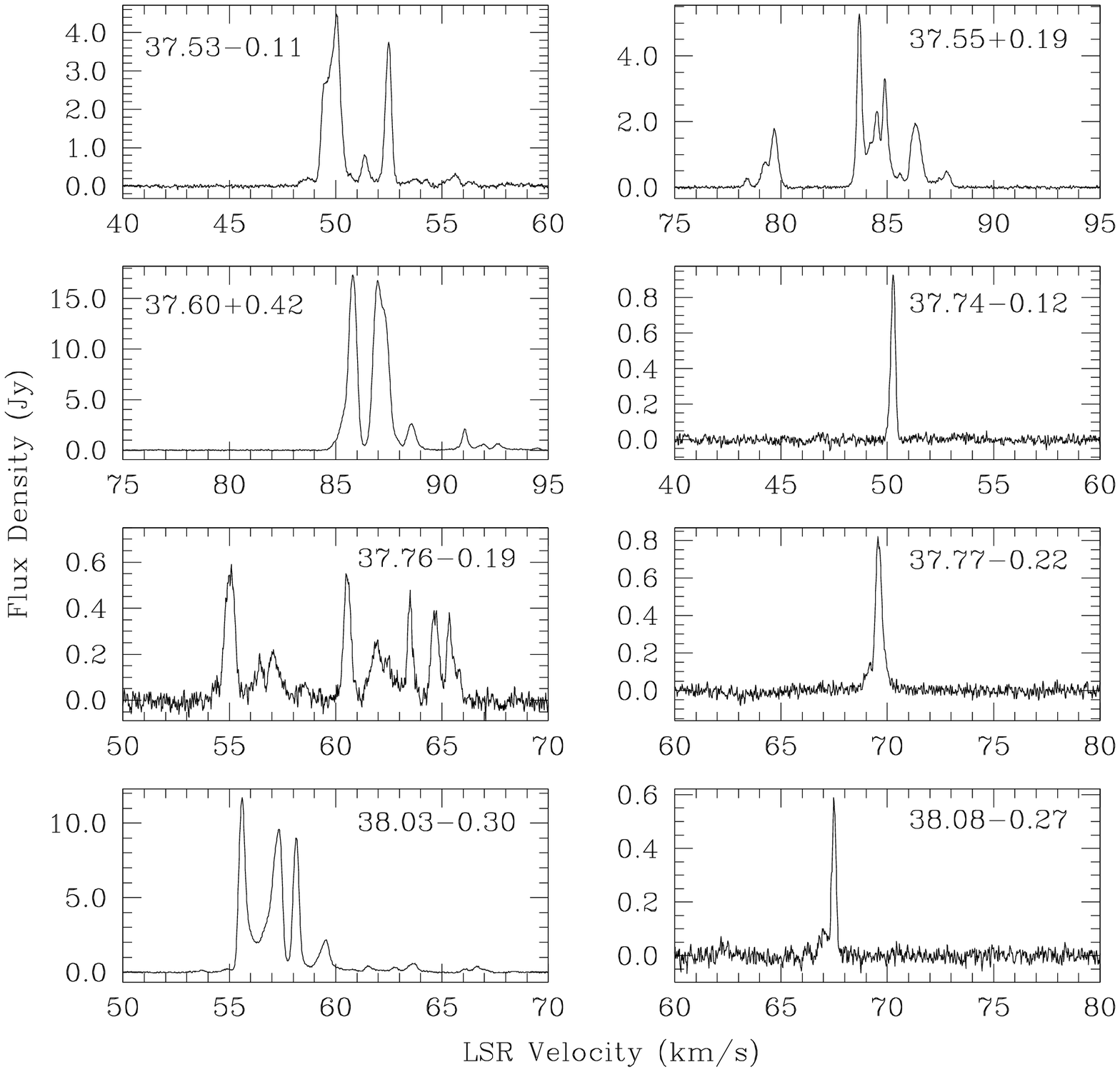}}\\[5mm]
\centerline{Fig. 6. --- Continued.}
\clearpage
{\plotone{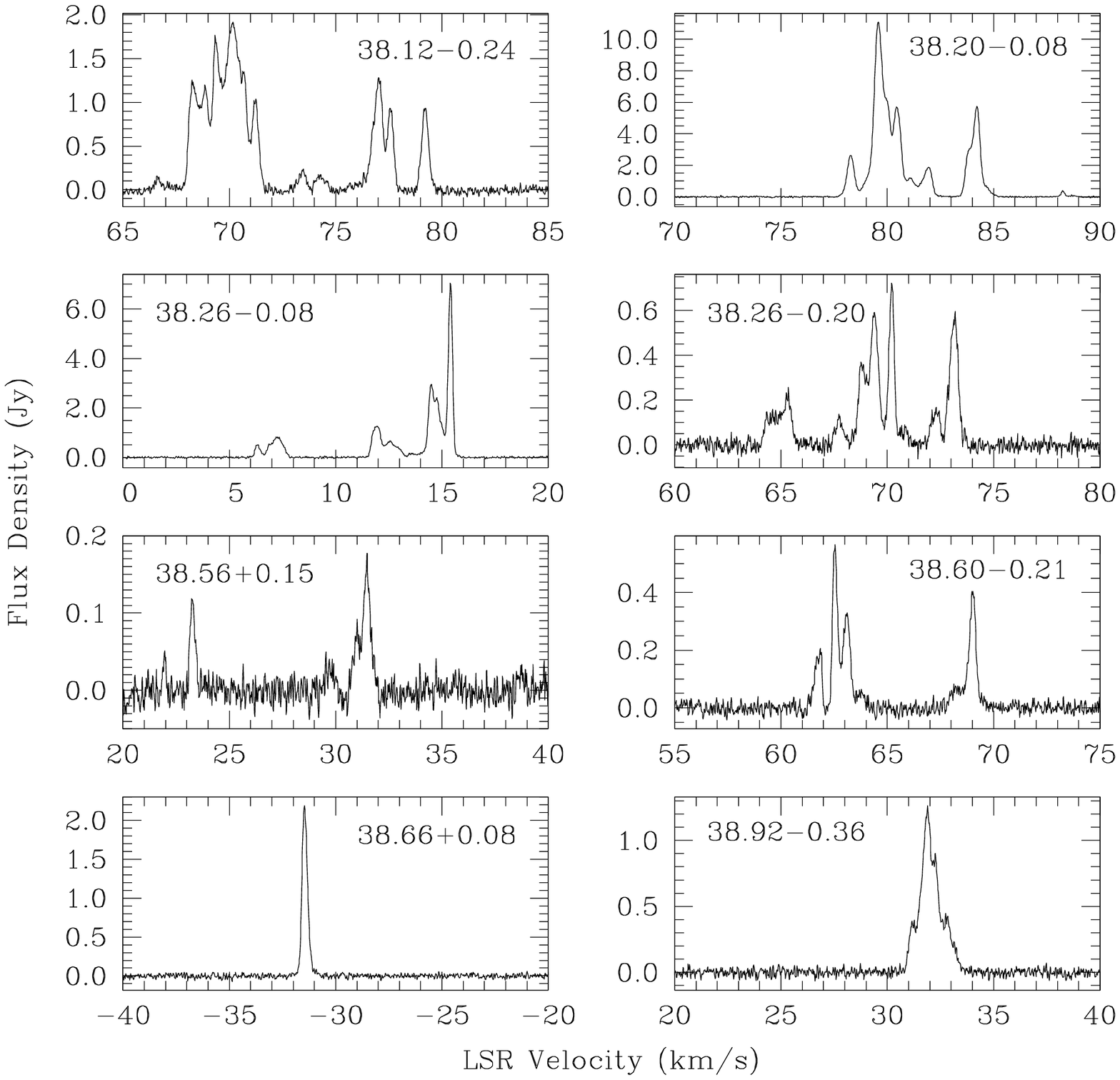}}\\[5mm]
\centerline{Fig. 6. --- Continued.}
\clearpage
{\plotone{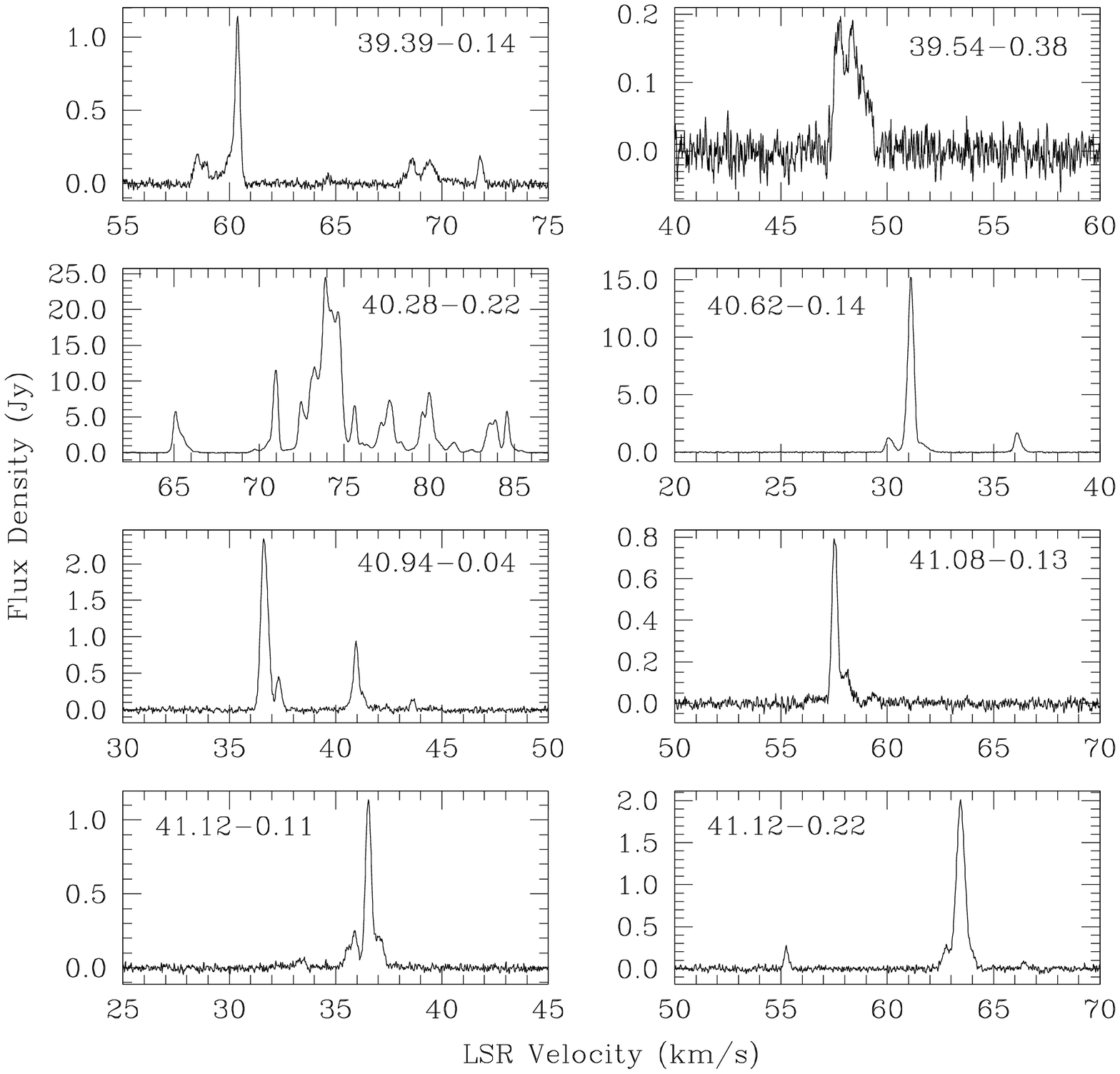}}\\[5mm]
\centerline{Fig. 6. --- Continued.}
\clearpage
{\plotone{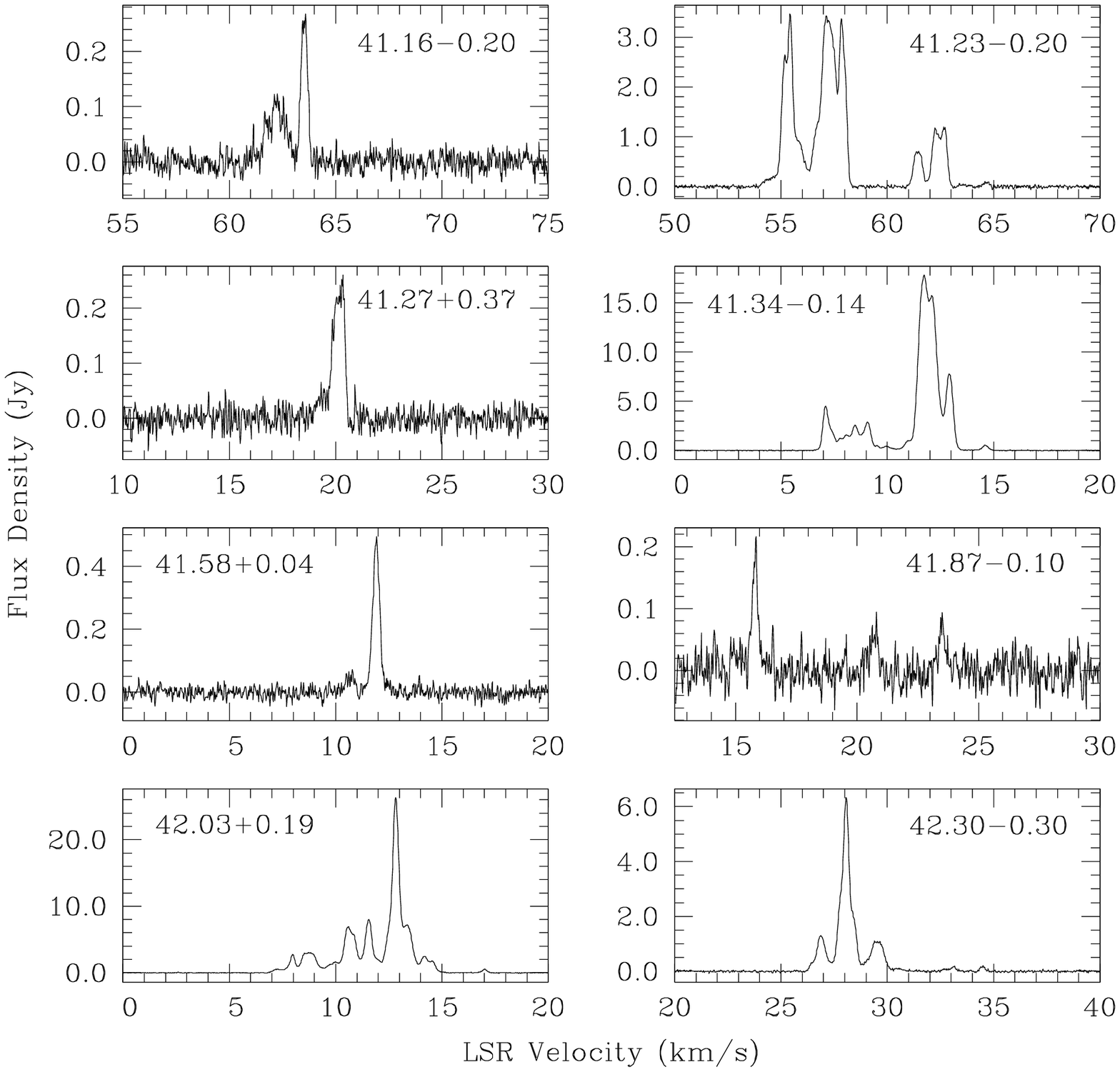}}\\[5mm]
\centerline{Fig. 6. --- Continued.}
\clearpage
{\plotone{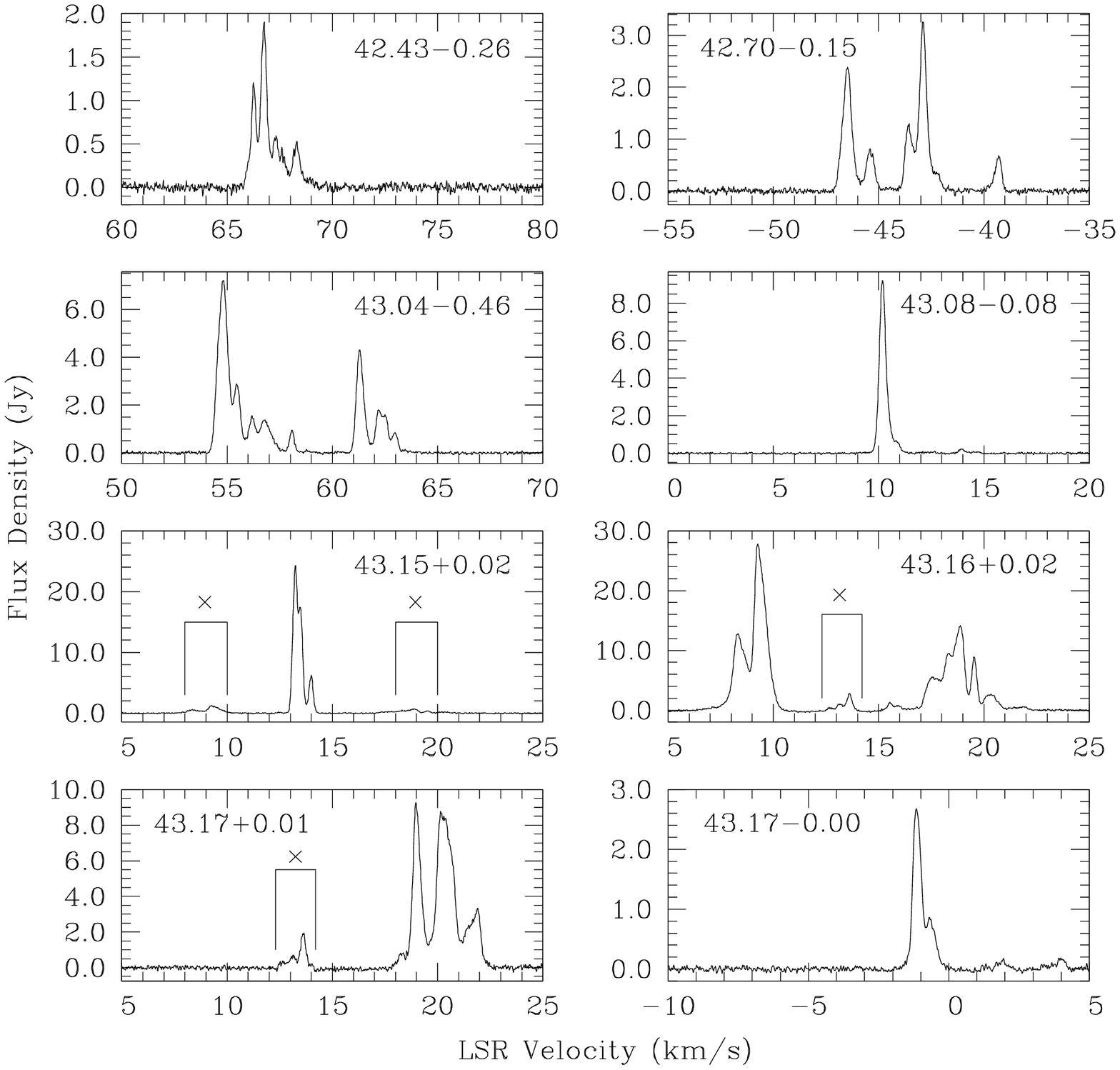}}\\[5mm]
\centerline{Fig. 6. --- Continued.}
\clearpage
{\plotone{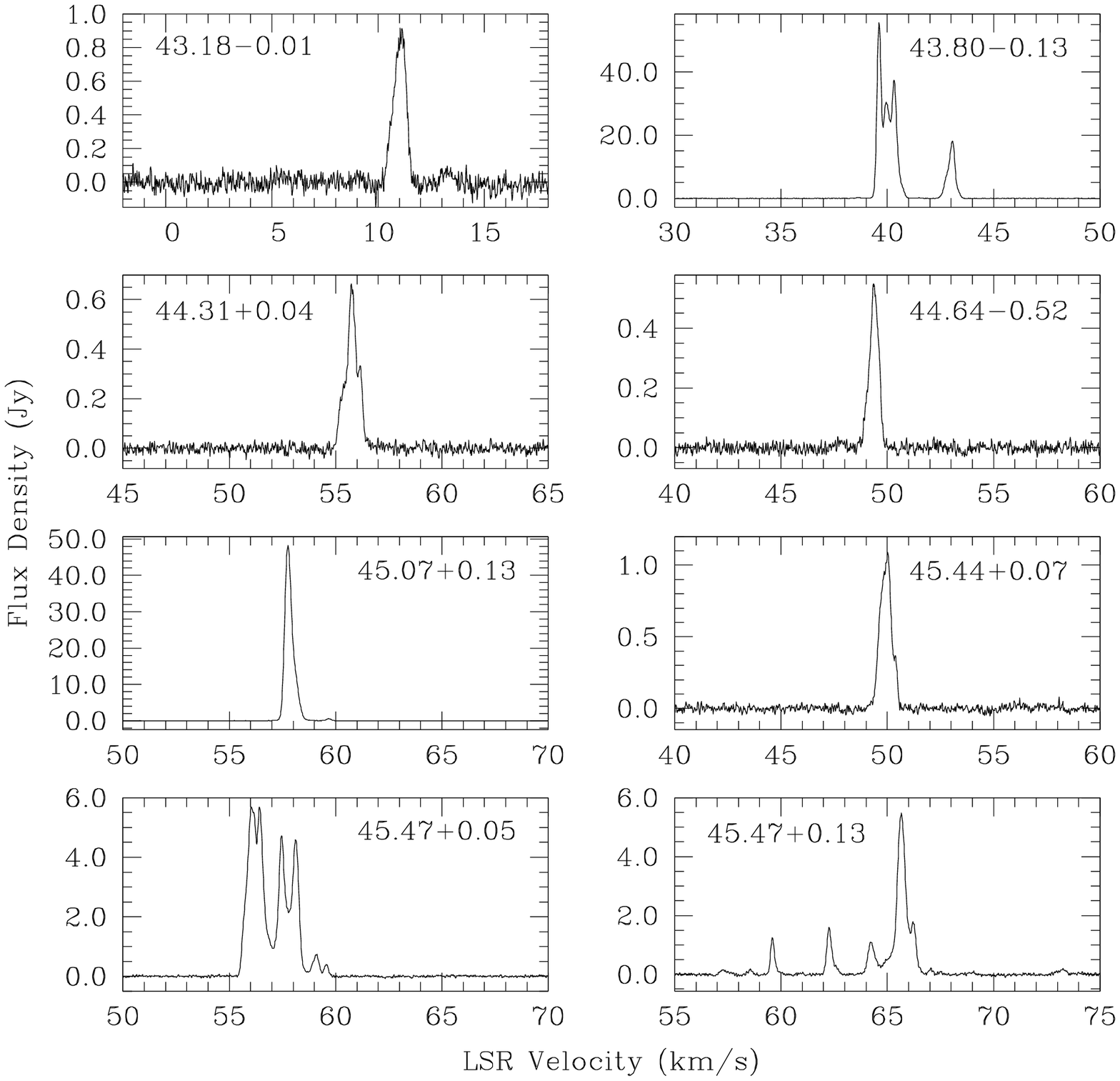}}\\[5mm]
\centerline{Fig. 6. --- Continued.}
\clearpage
{\plotone{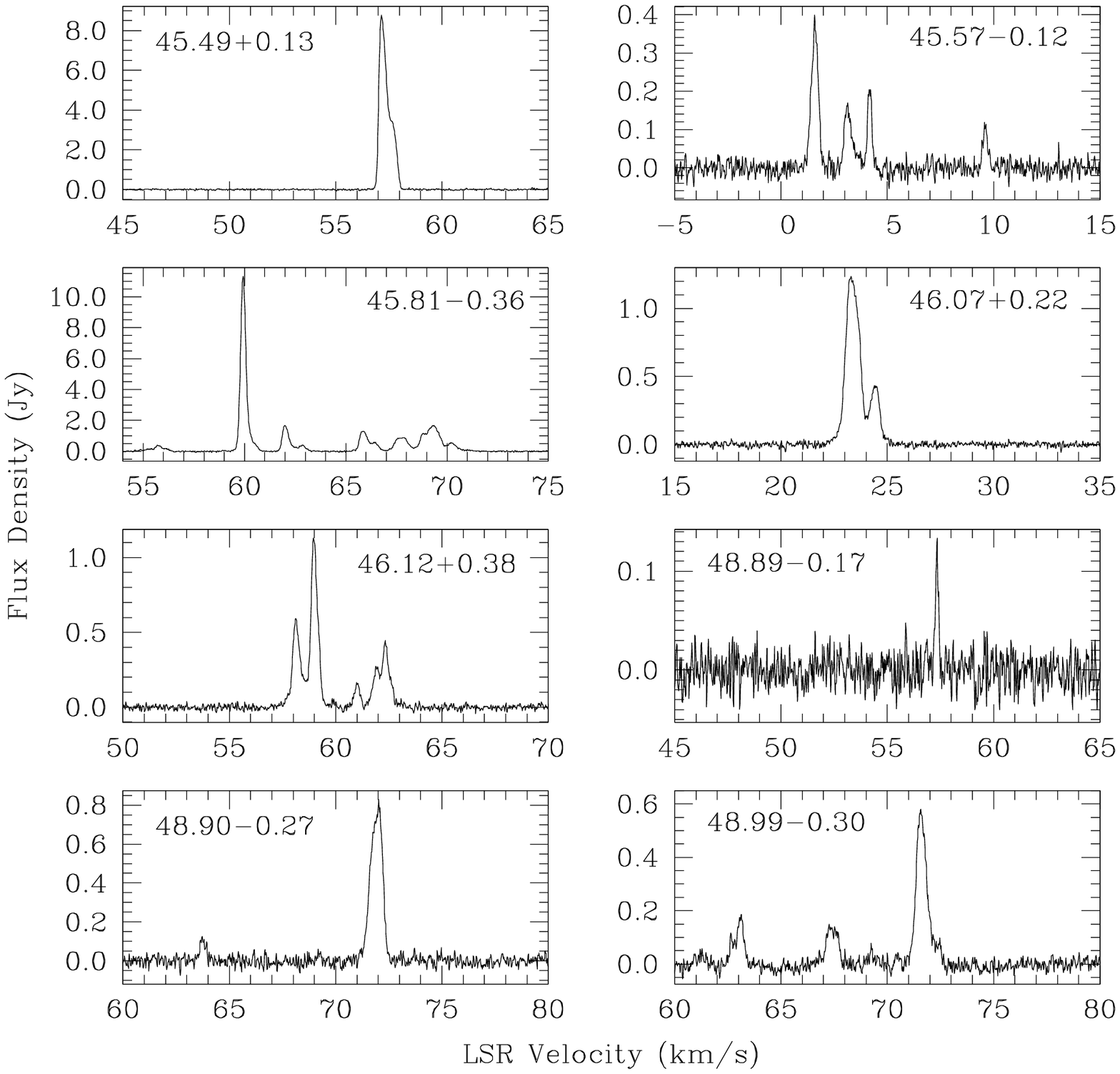}}\\[5mm]
\centerline{Fig. 6. --- Continued.}
\clearpage
{\plotone{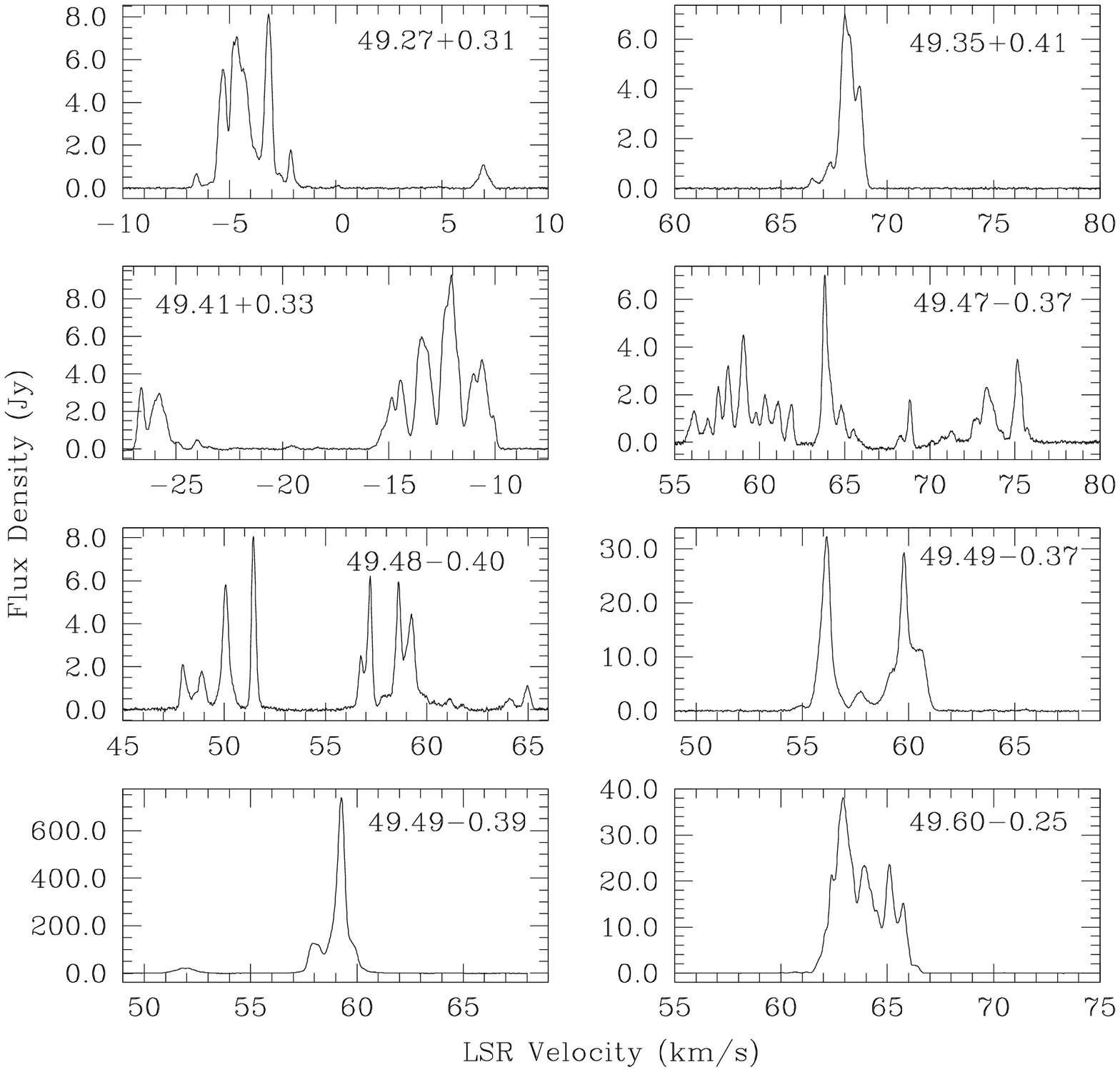}}\\[5mm]
\centerline{Fig. 6. --- Continued.}
\clearpage
{\plotone{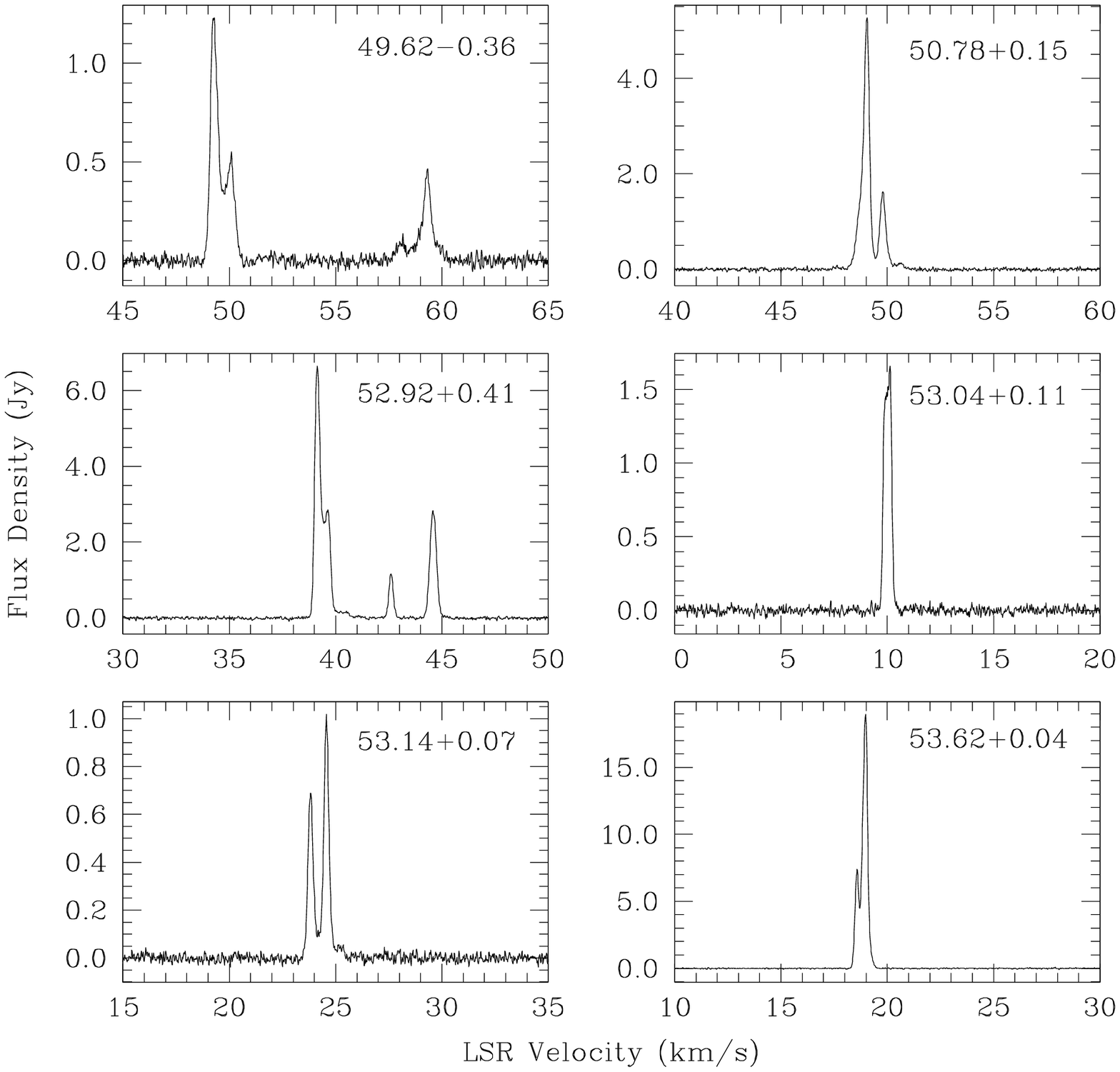}}\\[5mm]
\centerline{Fig. 6. --- Continued.}

\clearpage

\end{document}